\documentclass[a4paper,12pt]{article}

\usepackage[utf8]{inputenc}
\usepackage[T1]{fontenc}
\usepackage[english]{babel}
\usepackage{amsfonts,amsbsy,bm,euscript,mathrsfs}
\usepackage[tbtags]{amsmath}
\usepackage{authblk}
\usepackage{amssymb,stmaryrd,faktor}
\usepackage{graphicx}
\usepackage{caption}
\usepackage[title,titletoc]{appendix}
\usepackage{titlesec}
\usepackage[bookmarks=true,colorlinks=true,linkcolor=black,citecolor=blue,urlcolor=blue,bookmarksnumbered]{hyperref}
\usepackage{overpic}
\usepackage{upgreek}
\usepackage{mathabx} 
\usepackage{soul}
\usepackage[dvipsnames]{xcolor}

\usepackage[
    backend=bibtex,
    style=alphabetic,
maxbibnames=99,
giveninits=true,
minalphanames=1,
maxalphanames=3,url=false
  ]{biblatex}

  \bibliography{SDSM}
  \usepackage{setspace}

\usepackage{dsfont}
\usepackage{collref}
\usepackage{lmodern}
\usepackage{mathrsfs}
\usepackage{mathtools}
\usepackage{fixmath}
\usepackage{cancel}
\usepackage{cases}

\usepackage{bbm}
\usepackage{braket}
\usepackage{slashed}

\usepackage{booktabs}
\usepackage{subfig}
\usepackage{tikz}
\usetikzlibrary{plotmarks,calc,decorations,decorations.markings,decorations.pathmorphing}
\usetikzlibrary{shapes,arrows.meta,automata,positioning}
\usepackage{tikz-3dplot}
\usepackage{circuitikz}
\usepackage{slashed}

\usepackage[usenames,dvipsnames]{pstricks}
\usepackage{pstricks-add}
\usepackage{epsfig}
\usepackage{pst-grad} 
\usepackage{pst-plot} 
\usepackage[space]{grffile} 
\usepackage{etoolbox} 
\makeatletter 
\patchcmd\Gread@eps{\@inputcheck#1 }{\@inputcheck"#1"\relax}{}{}
\makeatother

\usepackage{blkarray}
\usepackage{comment}

\renewbibmacro{in:}{}

\newbibmacro{string+doi}[1]{%
  \iffieldundef{doi}{#1}{\href{http://dx.doi.org/\thefield{doi}}{#1}}}
\DeclareFieldFormat{title}{\usebibmacro{string+doi}{\mkbibemph{#1}}}
\DeclareFieldFormat[article]{title}{\usebibmacro{string+doi}{\mkbibquote{#1}}}

\numberwithin{equation}{section}

\usepackage{accents}
\newcommand\thickbar[1]{\accentset{\rule{.7em}{.8pt}}{#1}}
\newcommand\thickbarsmall[1]{\accentset{\rule{.5em}{.8pt}}{#1}}

\makeatletter
\renewcommand\section{\@startsection {section}{1}{\z@}
	{-3.5ex \@plus -1ex \@minus -.2ex}
	{2.3ex \@plus.2ex}
	{\normalfont\Large\bfseries}}
\renewcommand\subsection{\@startsection{subsection}{2}{\z@}
	{-3.25ex\@plus -1ex \@minus -.2ex}
	{1.5ex \@plus.2ex}
	{\normalfont\large\bfseries}}
\makeatother

\usepackage{titlesec}

\titleformat*{\section}{\large\bfseries\sf}
\titleformat*{\subsection}{\normalsize\bfseries}
\titleformat*{\subsubsection}{\normalsize\bfseries}
\titleformat*{\paragraph}{\large\bfseries}
\titleformat*{\subparagraph}{\large\bfseries}
\titlespacing{\author}{-5pt}{-5pt}{-5pt}[-5pt]

\newcommand{\dd}{\partial}

\newcommand{\CP}{\mathbb{CP}}
\newcommand{\CC}{\mathbb{C}}

\renewcommand{\tilde}{\widetilde}

\newcommand{\bea}{\begin{equation}}
\newcommand{\eea}{\end{equation}}
\newcommand{\bear}{\begin{eqnarray}}
\newcommand{\eear}{\end{eqnarray}}
\newcommand{\bearr}{\begin{eqnarray*}}
\newcommand{\eearr}{\end{eqnarray*}}

\newcommand{\tx}[1]{\text {#1}}
\newcommand{\cl}[1]{\mathcal {#1}}
\newcommand{\sr}[1]{\mathscr {#1}}
\newcommand{\bb}[1]{\mathbb {#1}}
\newcommand{\fk}[1]{\mathfrak {#1}}

\DeclareFontFamily{U}{solomos}{}
\DeclareErrorFont{U}{solomos}{m}{n}{10}
\DeclareFontShape{U}{solomos}{m}{n}{
  <-> s*[1.1]  gsolomos8r
}{}
   
\newcommand{\vkappa}{\text{\usefont{U}{solomos}{m}{n}\symbol{'153}}}

\newcommand{\nd}{\noindent}

\makeatletter
\newsavebox\BoxA
\newsavebox\BoxB
\newlength\LengthA

\newcommand*\obr[2][0.75]{%
	\sbox{\BoxA}{$\m@th#2$}%
	\setbox\BoxB\null
	\ht\BoxB=\ht\BoxA%
	\dp\BoxB=\dp\BoxA%
	\wd\BoxB=#1\wd\BoxA
	\sbox\BoxB{$\m@th\overline{\copy\BoxB}$}
	\setlength\LengthA{\the\wd\BoxA}
	\addtolength\LengthA{-\the\wd\BoxB}%
	\ifdim\wd\BoxB<\wd\BoxA%
	\rlap{\hskip 0.5\LengthA\usebox\BoxB}{\usebox\BoxA}%
	\else
	\hskip -0.5\LengthA\rlap{\usebox\BoxA}{\hskip 0.5\LengthA\usebox\BoxB}%
	\fi}
\makeatother

\begin{document}
	\title{Supersymmetric deformation of the $ \bb{CP}^{1} $ model \\ and its conformal limits}
	\author[1,2]{Dmitri Bykov\footnote{bykov@mi-ras.ru, dmitri.v.bykov@gmail.com}}
	\author[1,3]{Anton Pribytok\footnote{antons.pribitoks@pd.infn.it, a.pribytok@gmail.com}}
	\affil[1]{\small Steklov Mathematical Institute of Russian Academy of Sciences, \protect\\ Gubkina str. 8, 119991 Moscow, Russia}
	\affil[2]{\small Institute for Theoretical and Mathematical Physics,  \protect\\ Lomonosov Moscow State University,  119991 Moscow, Russia}
	\affil[3]{\small Dipartimento di Fisica e Astronomia Galileo Galilei, Università di Padova, \protect\\
	Via Marzolo 8, 35131, Padova, Italy}
	\date{}

\maketitle

\abstract{\nd We prove that the supersymmetric deformed $\CP^1$ sigma model (the generalization of the Fateev-Onofri-Zamolodchikov model) admits an equivalent description as a generalized Gross-Neveu model. This formalism is useful for the study of renormalization properties and particularly for calculation of the one- and two-loop $\beta$-function. We show that in the UV the superdeformed model flows to the super-Thirring CFT, for which we also develop a superspace approach. It is then demonstrated that the super-Thirring model is equivalent to a sigma model with the cylinder $\mathbb{R}\times S^1$ target space by an explicit computation of the correlation functions on both sides. Apart from that, we observe that the original model has another interesting conformal limit, given by the supercigar model, which as well could be described in the Gross-Neveu approach.} 

	\newpage
	\tableofcontents

	\newpage
	\section{Sigma models as Gross-Neveu models}
	
	In the present paper we continue investigation of a curious relation between 2D integrable sigma models and generalized Gross-Neveu (GN) type models with mixed bosonic/fermionic field content. This relation was proposed in~\cite{BykovHermitian, BykovSUSY, BykovGN, BykovRiemann} on the example of the $\CP^{n-1}$ sigma model. We refer the reader to those papers for an introduction to this subject.
	
	It is well established that many integrable sigma models admit trigonometric (and possibly elliptic) deformations (cf.~\cite{Klimcik0, Klimcik, DelducVicedo, Sfetsos} etc.). Perhaps one of the earliest and most well-known examples of a trigonometric deformation is the so-called \textit{sausage model} of \cite{FOZ} described by the following action\footnote{One could as well add a topological $\theta$-term, but we will not discuss it in the present paper.}:
	\begin{equation}\label{FOZ1}
		S_{\tx{FOZ}} = \int \; \frac{\dd X \thickbarsmall{\dd} X + \dd Y \thickbarsmall{\dd} Y }{a(t)+b(t) \cosh 2 Y} \,d^{2}z \;. 
	\end{equation}
The parameters $a>0$ and $b>0$ are the only ones to get renormalized at one loop, so that, remarkably, the RG equation\footnote{We choose the time direction $t\in (-\infty, 0)$ towards the IR.} ${dg_{ij} \over dt}=-{1\over 2\pi}R_{ij}$ reduces to a set of ODEs 
\bea\label{abRG}
\dot{a}={1\over 2\pi}b^2, \quad\quad\dot{b}={1\over 2\pi}ab\,.
\eea
In other words, despite not being homogeneous, the metric in~(\ref{FOZ1}) is stable under renormalization at one loop.

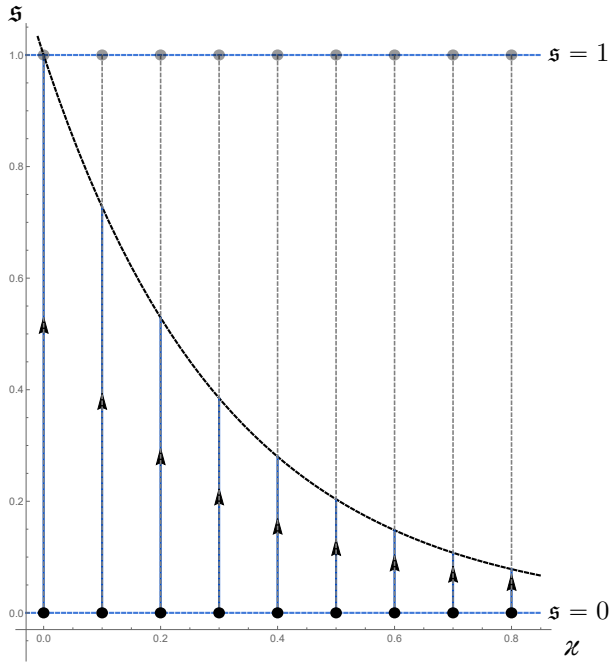
\begin{figure}[h]
	\centering
	\begin{overpic}[abs,scale=0.35,unit=1mm]{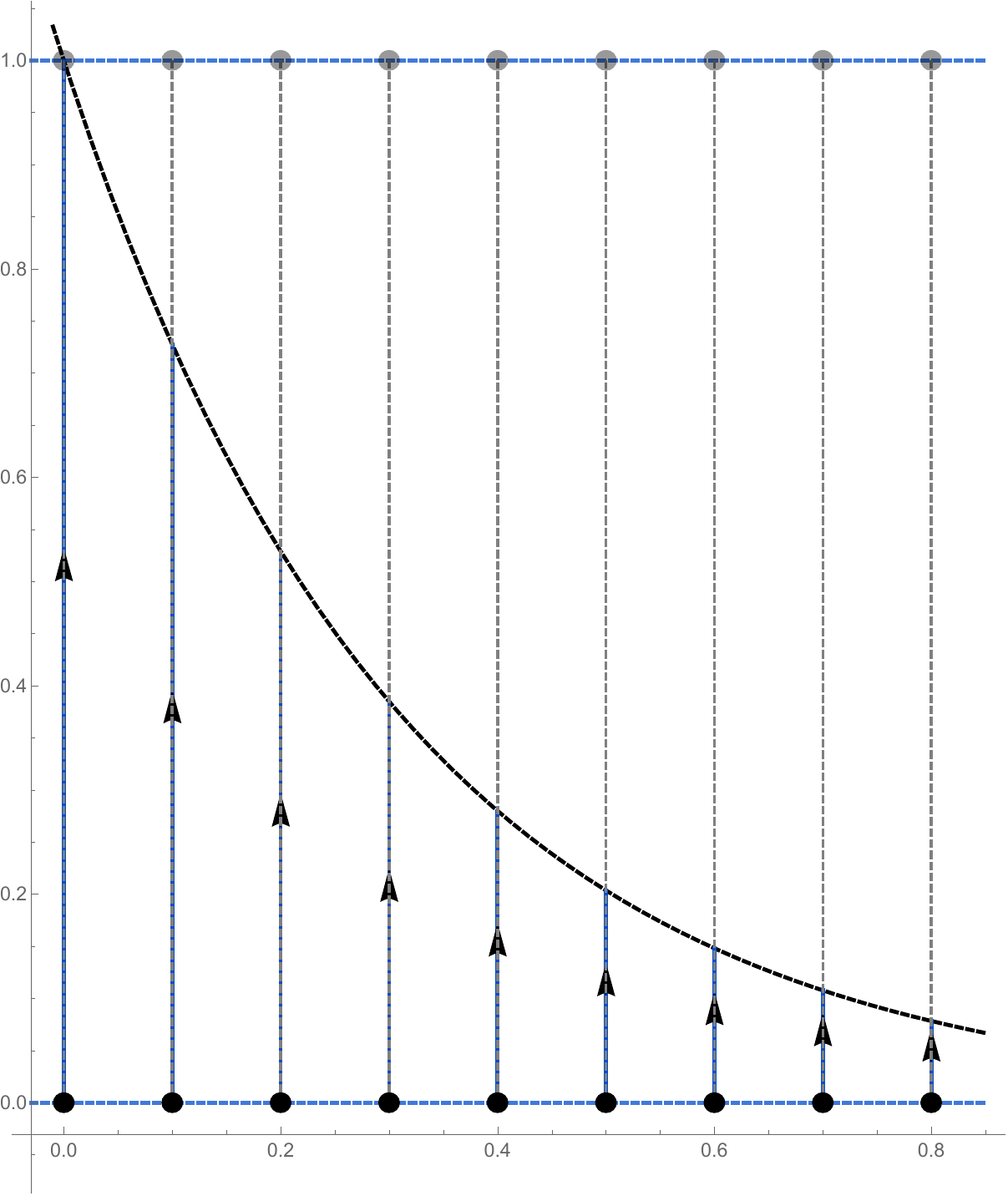}
		\put(0,85){$ \fk{s} $}
		\put(73,1){$ \vkappa $}
		\put(71,79.7){{\footnotesize $ \fk{s} = 1 $}}
		\put(71,5.7){{\footnotesize $ \fk{s} = 0 $}}
	\end{overpic}
	\caption{RG evolution trajectories in the $ (\vkappa,\fk{s}) $-plane flowing from UV to IR. The line $ \fk{s}=0 $ corresponds to the cylinder target space, whereas in the IR limit $ \fk{s} = 1 $ one recovers the spherical metric. The dashed curve indicates a time slice of the flow.}
	\label{RGFT}
\end{figure}

In fact, this theory is a one-parameter deformation of the $\CP^1$ sigma model with the metric\footnote{Recall that, in stereographic coordinates, the round metric has the form $ds^2\sim \frac{|du|^2}{\left(1+|u|^2\right)^2}$. }
	\begin{equation}\label{FOZmetric}
		ds^2=\frac{\fk{s}^{-{1\over 2}}-\fk{s}^{1\over 2}}{\vkappa}\,\frac{2\,|du|^2}{\left(\fk{s}^{1\over 2}+|u|^2\right)\left(\fk{s}^{-{1\over 2}}+|u|^2\right)}:=g_{u\thickbarsmall{u}}\,|du|^2
	\end{equation}
	that features two parameters, $\vkappa$ and $\fk{s}$. Clearly, $\vkappa$ is the overall scale, whereas $\fk{s}$ is the deformation parameter.  These are related to $a$ and $b$ via $ a = \frac{1+\fk{s} }{1 - \fk{s}}{\vkappa},\; b = \frac{2\sqrt{\fk{s}}}{1-\fk{s}} {\vkappa} $. In this new parametrization the RG equations~(\ref{abRG}) significantly simplify to 
	\bea\label{RGflow}
\dot{\fk{s}}={\vkappa\over \pi}\, \fk{s}\,,\quad\quad\dot{\vkappa}=0	\,,
	\eea
with the solution $\fk{s}=e^{{\vkappa\over \pi}t}$. The trajectories are plotted in~Fig.~\ref{RGFT}. In the IR limit $\fk{s}\to 1$ one recovers the standard metric on the 2-sphere in stereographic coordinates. The vanishing of the radius in this limit signals the onset of a strongly coupled regime typical of such~theories. In the opposite regime $\fk{s}\to 0$, in the UV, one has `asymptotic freedom', since the target space is a cylinder with the flat metric $ds^2\sim \frac{|du|^2}{|u|^2}$, and therefore the resulting sigma model is `quasi-free'. We will elaborate the details in section~\ref{Cylinder_CF} below. 

\vspace{0.3cm}
The model (\ref{FOZ1}) admits an extension with $\mathcal{N}=(2, 2)$ supersymmetry, since any two-dimensional geometry is K\"ahler. One of the goals of the present paper is to extend the correspondence between sigma models and Gross-Neveu models to the case of supersymmetric deformed models (e.g. the SUSY extension of the model~\eqref{FOZ1}), and to elaborate its consequences in the UV conformal limit. As we shall see in section~\ref{Susy_Def_Sec} below, the GN formulation of the deformed model explicitly involves the classical $r$-matrix. Moreover, $\mathcal{N}=(2, 2)$ SUSY invariance relies on a certain property of the $\mathfrak{sl}_2$ $r$-matrix. Generalisation of this construction to $\mathfrak{sl}_n$ with $n>2$ remains an open problem: presumably the difficulty here is related to the fact that, for $n>2$, the target space geometry of the SUSY model is of generalized K\"ahler type~\cite{Demulder, BykovLust} (except in the undeformed case, which was discussed in detail in~\cite{BykovSUSY}), while for $n=2$ it remains K\"ahler.

In section~\ref{inhomsec} we explicitly prove equivalence of the constructed GN model with the deformed $\CP^1$ sigma model in its geometric formulation by passing to inhomogeneous coordinates. We show that all geometric quantities that are present in the Lagrangian, such as Christoffel symbols and the Riemann tensor, automatically arise from the GN model upon integrating over the `momentum' variables ($v, \thickbarsmall{v}$). We also demonstrate explicitly that our construction produces interesting limits of the target space geometry that lead to SUSY versions of the  cylinder and cigar sigma models.
	
In section~\ref{betafuncsec} we proceed to compute the one- and two-loop $\beta$-function of the superdeformed GN model. In particular, we prove that the model is renormalizable up to two loops, in the sense that all divergences may be absorbed in the renormalization of the deformation parameter $\fk{s}$. The UV limit of the renormalization group flow corresponds to $\fk{s}\to 0$, where the GN model degenerates into a super-Thirring model -- a rather prominent CFT system that was first studied in the 80's~\cite{Freedman1, Freedman2} in the context of $bc$-$\beta\gamma$ ghost systems. 

To make SUSY manifest, in section~\ref{Super_Thirring_Superspace} we provide explicit superspace formulations of the super-Thirring model: curiously, the $\mathcal{N}=(2,2)$ formulation involves interesting combinations of chiral, twisted chiral and semi-chiral superfields in Euclidean space (in this setup the gauge field belongs to a semi-chiral multiplet). 

We then demonstrate that the super-Thirring model is equivalent to a sigma model with the cylinder $\mathbb{R}\times S^1$ target space. The relation is exact in quantum theory and may be checked at the level of correlation functions, which is what we do for the case of 2- and 4-point functions in section~\ref{Cylinder_CF}. On the side of the super-Thirring model the computation involves the summation of (crossed) ladder diagrams, whereas on the sigma model side it reduces to the computation of a correlation function of vertex operators. For arbitrary $n$-point functions the answer is given by a Koba-Nielsen type formula, as shown in section~\ref{npointsec}. 

In appendix~\ref{auxfieldsapp} one can find details on the elimination of auxiliary fields in ${\mathcal{N}=(1,1)}$ superspace. For completeness we also calculate the two-loop correction to the propagator of elementary fields in the Thirring model in appendix~\ref{2loopcorrapp}. This correction is present in the pure fermionic and pure bosonic Thirring models but vanishes in the supersymmetric case. Finally, in appendix~\ref{recurrsolapp} we solve the recursion relation for the $\ell$-loop contribution to the 4-point function in the super-Thirring model derived in Sec.~\ref{4ptsec}.

\section{Supersymmetric deformation}\label{Susy_Def_Sec}
	
	Supersymmetric formulations of the $\bb{CP}^{n-1}$ sigma model date back to~\cite{CremmerScherk, DAdda}. 
	In the present paper we address a class of \emph{deformed} supersymmetric sigma models, which admit a Gross-Neveu formulation. 
	
	More specifically, the supersymmetric GN model corresponding to the deformed $ \bb{CP}^{1} $ sigma model is defined  in terms of the fields
	\begin{equation}
		\begin{aligned}
			&  U_\alpha, \;V_\alpha  \qquad \alpha=1, 2\,, \qquad \tx{even} \quad \tx{(bosonic)}\,, \\ 
			&  B_\alpha, \; C_\alpha  \qquad \alpha=1, 2\,, \qquad \tx{odd} \quad \tx{(fermionic)}\,,
		\end{aligned}
	\end{equation}
	where $\alpha$ is an $\mathfrak{su}_2$-index (since $SU(2)$ is the symmetry group of the undeformed $ \bb{CP}^{1} $ model), which for brevity we will suppress henceforth. The action is\footnote{Taking the kinetic term imaginary is also conventional for mechanics in Euclidean time, since in that case  the path integral schematically takes the form $\int\,Dp\,Dq\,e^{-\mathcal{S}}=\int\,Dp\,Dq\,\mathrm{exp}\left(\int \left(i\,p \dot{q} -\mathsf{H}(p, q)\right)\,dt\right)$ with $\mathsf{H}(p, q)$ the Hamiltonian. One could equivalently view the coupling constant $\vkappa$ as being imaginary.}
	\bear \label{Def_Action}
	&&\mathcal{S}=\int\,i\,dz\!\wedge\! d\thickbarsmall{z}\;\sr{L}_{\fk{s}}\equiv 2\,\int\,d^2z\;\sr{L}_{\fk{s}}\,, \\ \label{Def_Lagrangian}
		&&\sr{L}_{\fk{s}} = \left(\sr{V}\thickbarsmall{\sr{D}}\sr{U} - \mathrm{c.c.}\right)
		+ {\vkappa\over 2} \, \tx{Tr} \left[ r_{\fk{s}}(J) \, \thickbarsmall{J} \,\right]
	\eear
	where $ \sr{U} $, $ \sr{V} $ form superdoublets of the $ U, \, V $ and $ B, \, C $ fields\footnote{In our conventions, complex conjugation acts on Grassmann variables according to the rule $\thickbarsmall{bc}\equiv \thickbarsmall{b}\thickbarsmall{c}$.},  and the current $ J $ is bilinear in those fields:   
	
	\begin{equation}\label{UVMu}
			\sr{U} = 
			\begin{pmatrix}
				U \\
				C
			\end{pmatrix} 
			\qquad
			\sr{V} = 
			\begin{pmatrix}
				V & B
			\end{pmatrix}
	\end{equation}
	\begin{equation}\label{currdef}
		J = U \otimes V - C \otimes B \, , \qquad \thickbarsmall{J} = \thickbarsmall{V} \otimes \thickbarsmall{U} + \thickbarsmall{B} \otimes \thickbarsmall{C}\,,
	\end{equation}
	where bar implies Hermitian conjugation. The kinetic part of the Lagrangian may be thought of as defining a $ pdq $-type form on the phase space of the model. In this case, $J$ is the moment map for the action of a global symplectomorphism group. Throughout this paper, though, we will simply interpret $J$ as a Kac-Moody current of a free $\beta\gamma$-system.
	
	The covariant superderivative introduces a gauge superfield $ \thickbarsmall{\fk{A}}_{\tx{super}} $ that includes the component gauge fields $ \thickbarsmall{A} $ and $ \thickbar{W} $:
	\begin{equation}\label{supergaugefields}
		\thickbarsmall{\sr{D}} = \thickbarsmall{\partial} +
		i\, \thickbarsmall{\fk{A}}_{\tx{super}}
		\qquad \quad
		\thickbarsmall{\fk{A}}_{\tx{super}} = 
		\begin{pmatrix}
			\thickbarsmall{A} & 0 \\
			\thickbar{W} & \thickbarsmall{A}
		\end{pmatrix} \, .
	\end{equation}
	Variation of the action w.r.t. the gauge fields $\thickbarsmall{A}$ and $\thickbar{W}$ provides the first class constraints
	\begin{equation}\label{SConstraints-1}
		\begin{cases}
		\thickbarsmall{A}: \quad \fk{C}_{2} = V \cdot U + B \cdot C = 0 \\
			\thickbar{W}: \quad \fk{C}_{1} = B \cdot U = 0 \,,
		\end{cases}
	\end{equation}
	which will be exploited in the foregoing discussion.
	
	The action of the $ r_{\fk{s}} $ operator on a $2\times 2$-matrix $\cl{O}$ can be defined as follows:
		\begin{eqnarray} \label{rmatrixdef}
			r_{\fk{s}} \left[ \cl{O} \right] = \dfrac{1}{1-\fk{s}}
			\begin{pmatrix}
				\frac{1}{2} (\fk{s} + 1) \cdot \cl{O}_{11} & \sqrt{\fk{s}} \cdot \cl{O}_{12} \\
				\sqrt{\fk{s}} \cdot \cl{O}_{21} & \frac{1}{2} (\fk{s} + 1) \cdot \cl{O}_{22}
			\end{pmatrix}
			\quad ,
		\end{eqnarray}
where $\fk{s}$ is a deformation parameter. Obviously $r_{\fk{s}}$ is symmetric w.r.t. the quadratic form given by the trace: $\mathrm{Tr}(\cl{O}_2 \, r_{\fk{s}} \left[ \cl{O}_1 \right])=\mathrm{Tr}(\cl{O}_1 \, r_{\fk{s}} \left[ \cl{O}_2 \right])$.

	It is crucial to note that  superinvariance of the action following from Lagrangian (\ref{Def_Lagrangian}) is not manifest and hence must be proven separately. To this end let us consider worldsheet supertransformations~\cite{Friedan:1985sit,FMS} of the form (here $\epsilon_{1,2}$ are complex Grassmann parameters) 
	\begin{equation}\label{Supersymmetry_1}
		\delta U=\epsilon_1 C, \quad \delta B=-\epsilon_1 V, \quad \delta C=-\epsilon_2 \, \partial U, \quad \delta V=\epsilon_2 \, \partial B 
	\end{equation}
	generated by\footnote{In the literature the supercharges are often denoted as $Q_1=Q, Q_2=\thickbarsmall{Q}$, however they are only complex conjugate in Minkowski signature. Since we are dealing with Euclidean worldsheet signature throughout this paper, we will avoid this notation.}
	\begin{equation}\label{Supersymmetry_2}
		\delta = \epsilon_{1} Q_{1} + \epsilon_{2} Q_{2} \, , \qquad Q_{1}^{\,2} = Q_{2}^{\,2} = 0 \, , \qquad \{ Q_1 , Q_2 \} = \partial \, , 
	\end{equation}
	which, along with the antiholomorphic transformations, complete the $ \cl{N} = (2,2) $ supersymmetry algebra. Next one can compute the field variations of the Lagrangian (\ref{Def_Lagrangian}), which leads to the EOM\footnote{Very explicitly, the $\CC^\ast$-covariant derivatives are defined as follows:
		\begin{equation*}
			\thickbarsmall{D} U = \thickbarsmall{\partial}U + i \thickbarsmall{A} U \,,\quad  \thickbarsmall{D} V = \thickbarsmall{\partial}V -i \thickbarsmall{A} V\,,\quad 
			\thickbarsmall{D} B = \thickbarsmall{\partial} B - i \thickbarsmall{A} B\,,\quad \thickbarsmall{D} C = \thickbarsmall{\partial} C + i \thickbarsmall{A} C\,,
		\end{equation*}
together with the complex conjugate expressions.}
	\begin{equation}\label{sL_EOM_Holomorphic}
		\begin{array}{l}
			\delta_{U}: \thickbarsmall{D}V + i\thickbar{W} B - {\vkappa\over 2} V r_{\fk{s}}[\thickbarsmall{J}\,] = 0 \\ 
			\delta_{V}: \thickbarsmall{D} U + {\vkappa\over 2} \,r_{\fk{s}}[\thickbarsmall{J}\,] U  = 0
		\end{array}
		\qquad
		\begin{array}{l}
			\delta_{B}: \thickbarsmall{D} C + i\thickbar{W} U + {\vkappa\over 2}\, r_{\fk{s}}[\thickbarsmall{J}\,] C  = 0 \\ 
			\delta_{C}: \thickbarsmall{D} B - {\vkappa\over 2} B r_{\fk{s}}[\thickbarsmall{J}\,] = 0 \,,
		\end{array}
	\end{equation}
	with analogous equations for the antiholomorphic sector. By exploiting (\ref{sL_EOM_Holomorphic}) one can derive EOM for the currents: 
	\begin{equation}\label{MM_EOM}
		\boxed{ \qquad \thickbarsmall{\partial} J = {\vkappa\over 2}\, [J, r_{\fk{s}}[\thickbarsmall{J}\,]] \,,\qquad \partial\thickbarsmall{J} = -{\vkappa\over 2}\, \left[ r_{\fk{s}}[J], \thickbarsmall{J}\, \right] \qquad}
	\end{equation} 
	To check supersymmetry of the above system~(\ref{Def_Lagrangian}) one should first compute the variation of the current $J$ 
	\begin{equation}\label{hatmudef}
		\delta J = \epsilon_{2} \partial \hat{J} \, , \qquad \hat{J} = U \otimes B
	\end{equation}
	Variation of the interaction term leads us to a characteristic commutator of the form 
	\begin{equation}\label{dInt}
		\begin{aligned}
			{\vkappa\over 2} \, \delta \left( \tx{Tr}\left( J \, r_{\fk{s}}[\thickbarsmall{J}] \right) \right) \approx -{\vkappa\over 2} \epsilon_{2} \tx{Tr} \left( \hat{J} \, r_{\fk{s}}[\partial\thickbarsmall{J}] \right) 
			= \left({\vkappa\over 2}\right)^{2} \epsilon_{2} \tx{Tr}\left( [r_{\fk{s}}[\hat{J}] , r_{\fk{s}}[J]] \thickbarsmall{J} \right)
		\end{aligned}
	\end{equation} 
	where we have applied integration by parts and cyclicity of the trace (analogous for the antiholomorphic counterpart). We then use equations of motion \eqref{MM_EOM} along with symmetry of the $r$-matrix. In order to simplify the above equation we will derive the following remarkable property of the $r$-matrix~(\ref{rmatrixdef}):
	\bea\label{rmatalgsu2_0}
	\dot{r}_{\fk{s}}([\mathbb{A}, \mathbb{B}])=[r_{\fk{s}}(\mathbb{A}), r_{\fk{s}}(\mathbb{B})]\,,
	\eea
	where $\mathbb{A}$ and $\mathbb{B}$ are two arbitrary matrices. Here, by definition, $\dot{r}_{\fk{s}}:=\fk{s} {d r_{\fk{s}}\over d \fk{s}}$. This identity may be checked by an explicit calculation, and we will discuss its importance in the context of renormalization properties of the model in section~\ref{betafuncsec}.  Substituting $\mathbb{A}=\hat{J}$ and $\mathbb{B}=J$ and using the identity
	\bea\label{mu_vcommutator}
	[\hat{J} , J]=\fk{C}_{1}\,J-\fk{C}_{2}\,\hat{J}\,
	\eea
	that follows directly from the definitions~(\ref{UVMu}) and (\ref{hatmudef}), we find from~(\ref{rmatalgsu2_0}):
	\bea\label{rmu_commutator}
	[r_{\fk{s}}(\hat{J}), r_{\fk{s}}(J)]=\fk{C}_{1}\,\dot{r}_{\fk{s}}(J)-\fk{C}_{2}\,\dot{r}_{\fk{s}}(\hat{J})\,.
	\eea
	As a result, the r.h.s. of~(\ref{dInt}) vanishes modulo the constraints, thus proving supersymmetric invariance of the model \eqref{Def_Lagrangian}. 
	
	It is important to note that~\eqref{rmatalgsu2_0} does not hold for generic $n$, so that a separate treatment of deformed $\bb{CP}^{n-1}$ is necessary for $n>2$. A related fact is that for $ n>2 $ the deformed geometry becomes generalized Kähler~ \cite{Demulder:2019vvh,Demulder, BykovLust} and is yet to be embedded in our formalism. For the $n=2$ case these issues do not arise, and in the next section we shall explicitly prove equivalence of (\ref{Def_Lagrangian}) with the deformed super-$\bb{CP}^{1}$ model (providing an extension of the pure bosonic model of~\cite{FOZ}).

	\section{Equivalence with the $\CP^1$ sigma model}\label{inhomsec}
	
	Having proven that the model has $\mathcal{N}=(2, 2)$ supersymmetry, one can now further proceed to study its properties and geometric interpretation. We shall start from the Lagrangian~(\ref{Def_Lagrangian}) of the previous section. It follows from~(\ref{supergaugefields}) that gauge symmetry acts on the fields as
	\bea
	U \mapsto \uplambda U\,,\quad\quad C \mapsto \uplambda C+\upxi U\,,
	\eea
	where $\uplambda \in \CC^\ast$ and $\upxi \in \CC_{F}$ (the Grassmann vector space). In order to compare to the geometric formulation of the sigma model, it is useful to use inhomogeneous coordinates, which is nothing but a special way of fixing the gauge symmetry. This amounts to setting 
	\begin{equation}
		U_{1} = 1,  \quad\quad C_{1} = 0
	\end{equation}
	For convenience we will also redefine $U_{2} = u, \;V_2=v,\; C_2=c,\;B_2=b$. We may now solve the constraints~(\ref{SConstraints-1}), 
	which results in 
	\begin{equation}
		\begin{aligned}
			U = 
			\begin{pmatrix}
				1 \\
				u
			\end{pmatrix} \,,
			\quad
			V = 
			\begin{pmatrix}
				-b c -u v &
				v
			\end{pmatrix}\,,
			\quad
			C = 
			\begin{pmatrix}
				0 \\
				c
			\end{pmatrix} \,,\quad 
			B = b
			\begin{pmatrix}
				-u &
				1
			\end{pmatrix}\,,
		\end{aligned}
	\end{equation}
	and accordingly for the conjugate variables. The resulting Lagrangian in the inhomogeneous gauge takes the form
	\bear \label{GNinhom}
\sr{L}_{\fk{s}} = v \thickbar{\dd}u+b\thickbar{\dd}c+\thickbarsmall{u}\dd \thickbarsmall{v}-\thickbarsmall{c}\dd \thickbarsmall{b}+
\frac{\vkappa \,\fk{s}^{1\over 2}}{1-\fk{s}}\,\left[\upalpha |v|^2+\upbeta \left(vu\thickbarsmall{b}\thickbarsmall{c}+\thickbarsmall{v}\thickbarsmall{u} b c\right)+\upgamma b  c \thickbarsmall{b}\thickbarsmall{c}\right]\,, 
	\eear
	where 
	\bear
	&&\upalpha={1\over 2}\left(1+\left(\fk{s}^{1\over 2}+\fk{s}^{-{1\over 2}}\right)|u|^2+|u|^4\right)={1\over 2}\left(\fk{s}^{1\over 2}+|u|^2\right)\left(\fk{s}^{-{1\over 2}}+|u|^2\right)\,,\\
	&&\upbeta={1\over 2}\left(\fk{s}^{1\over 2}+\fk{s}^{-{1\over 2}}+2|u|^2\right)\,,\quad\quad \quad 
	\upgamma={1\over 2}\left(\fk{s}^{1\over 2}+\fk{s}^{-{1\over 2}}+4 |u|^2\right)\,.
	\eear
	Extremizing the action w.r.t. the momenta $v, \thickbarsmall{v}$, one gets
	\bea
	v_0={1\over \upalpha}\left(\frac{\fk{s}^{-{1\over 2}}-\fk{s}^{1\over 2}}{\vkappa} \dd \thickbarsmall{u}-\upbeta \thickbarsmall{u}\, bc\right)\,,\quad\quad \thickbarsmall{v}_{0}=-{1\over \upalpha}\left(\frac{\fk{s}^{-{1\over 2}}-\fk{s}^{1\over 2}}{\vkappa} \thickbar{\dd} u+\upbeta u\, \thickbarsmall{b}\thickbarsmall{c}\right)
	\eea
	Substituting these values back in the Lagrangian, one arrives at its final geometric form:
	\bear\label{lagrcomp}
	&&\sr{L}_{\fk{s}} =\frac{\fk{s}^{-{1\over 2}}-\fk{s}^{1\over 2}}{\vkappa}\,\frac{|\thickbar{\dd}u|^2}{\upalpha}+b \thickbar{D}c-\thickbarsmall{c}D \thickbarsmall{b}+\frac{\vkappa}{\fk{s}^{-{1\over 2}}-\fk{s}^{1\over 2}}\,\left(\upgamma-{\upbeta^2 \over \upalpha} |u|^2\right)\,b  c \thickbarsmall{b}\thickbarsmall{c}\,,\\
	&&\textrm{with}\quad\quad \thickbar{D}c=\thickbar{\dd}c-{\upbeta \over \upalpha}\,\thickbarsmall{u} \thickbar{\dd}u\,c\,,\quad\quad D\thickbarsmall{b}=\dd\thickbarsmall{b}+{\upbeta \over \upalpha}\,u \dd \thickbarsmall{u}\,\thickbarsmall{b}
	\eear
	Clearly, the first term encodes the metric~(\ref{FOZmetric}) of the deformed $\CP^1$ model. The coefficient functions appearing in the covariant derivatives nicely turn out to be exactly the Christoffel symbols of the K\"ahler geometry: $\Gamma^{u}_{uu}={\dd \log{g_{u\thickbarsmall{u}}}\over \dd u}=-{\upbeta \over \upalpha}\thickbarsmall{u}$\,, so that
	\bea
	\thickbar{D}c=\thickbar{\dd}c+ \Gamma^{u}_{uu} \thickbar{\dd}u\,c\,,\quad\quad D\thickbarsmall{b}=\dd\thickbarsmall{b}-\Gamma^{\thickbarsmall{u}}_{\thickbarsmall{u}\,\thickbarsmall{u}} \dd \thickbarsmall{u}\,\thickbarsmall{b}\,.
	\eea
	Finally, the coefficient of the quartic fermionic piece in~(\ref{lagrcomp}) is proportional to the Riemann tensor\footnote{Indeed,
	\begin{equation*}
		\frac{\vkappa}{\fk{s}^{-{1\over 2}}-\fk{s}^{1\over 2}}\,\left(\upgamma-{\upbeta^2 \over \upalpha} |u|^2\right)={1\over 2} \frac{\vkappa}{\fk{s}^{-{1\over 2}}-\fk{s}^{1\over 2}}\,\frac{\left(\fk{s}^{1\over 2}+\fk{s}^{-{1\over 2}}\right)\left(1+|u|^4\right)+4 |u|^2}{\left(\fk{s}^{1\over 2}+|u|^2\right)\left(\fk{s}^{-{1\over 2}}+|u|^2\right)}=g^{u\thickbarsmall{u}} g^{u\thickbarsmall{u}}R_{u\thickbarsmall{u}u\thickbarsmall{u}}\,.
	\end{equation*}
	}. 
	Altogether the Lagrangian~(\ref{lagrcomp}) has the standard form of an ${\mathcal{N}=(2, 2)}$ SUSY sigma model (cf.~\cite[Chapter 13]{MirrorBook}):
	\bea
	\sr{L}_{\fk{s}} =g_{u\thickbarsmall{u}}|\thickbar{\dd}u|^2+b \thickbar{D}c-\thickbarsmall{c}D \thickbarsmall{b}+R_{u\thickbarsmall{u}u\thickbarsmall{u}}\,g^{u\thickbarsmall{u}} g^{u\thickbarsmall{u}}\,b  c \thickbarsmall{b}\thickbarsmall{c}\,.
	\eea
	Thus, we have proven equivalence of the supersymmetrised GN model~(\ref{Def_Lagrangian}) with the deformed SUSY $\CP^1$ model.
	
\begin{figure}
\centering
\begin{overpic}[abs, unit=0.85mm,scale=0.85]{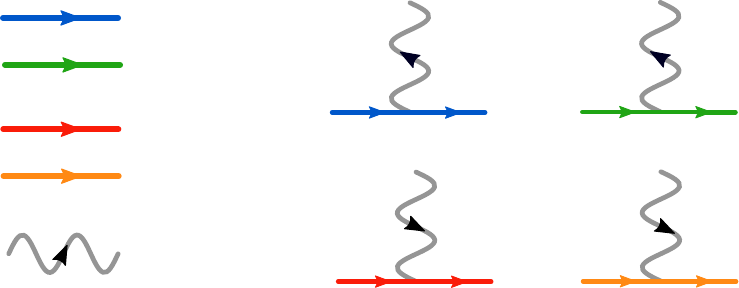}
    \put(65,45){$a$}
    \put(107,45){$a$}
    \put(65,17){$a$}
    \put(107,17){$a$}
    \put(76,38){$i\,\vkappa^{1\over 2} (r_{\fk{s}}(\tau_a))_{ij}$}
    \put(119,38){$i\,\vkappa^{1\over 2} (r_{\fk{s}}(\tau_a))_{ij}$}
    \put(77,8){$i\,\vkappa^{1\over 2} (\tau_a)_{ji}$}
    \put(119,8){$i\,\vkappa^{1\over 2} (\tau_a)_{ji}$}
    \put(-3,5){$a$}
    \put(22,5){$b$}
    \put(28,5){$\delta^{(2)}(z)\, \delta^{ab}$}
    \put(28,45){${1\over 2\pi z}\,\delta^{ij}$}
    \put(28,37){${1\over 2\pi z}\,\delta^{ij}$}
    \put(28,26){$-{1\over 2\pi \thickbarsmall{z}}\,\delta^{ij}$}
    \put(28,18){$-{1\over 2\pi \thickbarsmall{z}}\,\delta^{ij}$}
    \put(-3,45){$i$}
    \put(-3,37){$i$}
    \put(-3,26){$i$}
    \put(-3,18){$i$}
    \put(22,45){$j$}
    \put(22,37){$j$}
    \put(22,26){$j$}
    \put(22,18){$j$}
    \put(-20,45){$\langle U^i V^j\rangle$}
\put(-20,37){$\langle B^i C^j\rangle$}
\put(-20,26){$\langle \thickbar{U}^i \thickbar{V}^j\rangle$}
\put(-20,18){$\langle \thickbar{B}^i \thickbar{C}^j\rangle$}
\put(-20,5){$\langle \mathcal{B}^a \thickbar{\mathcal{B}}^b\rangle$}
\put(55,25){$i$}
\put(80,25){$j$}
\put(55,-4){$i$}
\put(81,-4){$j$}
\put(97,25){$i$}
\put(123,25){$j$}
\put(97,-4){$i$}
\put(123,-4){$j$}
\end{overpic}
\vspace{0.5cm}
\captionsetup{singlelinecheck=off}
\caption[Fig1]{Feynman rules of the deformed supersymmetric GN model~(\ref{Def_Lagrangian}):
\begin{itemize}
\item To simplify Feynman diagrams, we have introduced  `Hubbard-Stratonovich'-type fields $\mathcal{B}, \thickbar{\mathcal{B}}$ denoted by grey wavy lines, which split the quartic vertices into cubic ones. 
\item The matrices $\tau_a$ standing in the vertices are the generators of a Lie algebra $\mathfrak{g}_{\CC}$, which for most of the paper is taken to be $\mathfrak{sl}_2$. Nevertheless, it is useful to keep this more abstract notation to allow for certain generalizations below.
\end{itemize}
\par\noindent\rule{\textwidth}{0.5pt}
 }
\label{feynfig}
\end{figure}

\subsection{Scaling limits}

As in the bosonic case, there are two different interesting scaling limits of the resulting model. The first limit amounts to simply taking $\fk{s}\to 0$: here one gets
\bea
\sr{L}_{\mathrm{cyl}} ={2\over \vkappa}\,\frac{|\thickbar{\dd}u|^2}{|u|^2}+b \thickbar{D}c-\thickbarsmall{c}D \thickbarsmall{b}\,,
\eea
where the Christoffel symbols in the limit simplify to $\Gamma^{u}_{uu}=-{1\over u}$. This is a sigma model with target space the cylinder $\mathbb{R}\times S^1$. In fact, the change of variables
	\bea\label{fermCOV0}
c\mapsto u \,c, \quad b\mapsto u^{-1} \,b\,
	\eea
fully decouples the bosonic and fermionic variables, so that one arrives at the free theory 
\bea\label{cylfreetheory}
\sr{L}_{\mathrm{cyl}} ={2\over \vkappa}\,\frac{|\thickbar{\dd}u|^2}{|u|^2}+b \thickbar{\dd}c-\thickbarsmall{c}\dd \thickbarsmall{b}\,.
\eea
We will study it in great detail in section~\ref{Cylinder_CF} below. In the next section, by calculating the beta function of the deformed GN model, we will show that $\fk{s}\to 0$ is the conformal UV limit of the model.

The second limit involves first scaling the coordinate as  $u \to \fk{s}^{1\over 4} u$ and then taking the limit $\fk{s}\to 0$. As a result, one arrives at
\bea
\sr{L}_{\mathrm{cyl}} ={2\over \vkappa}\,\frac{|\thickbar{\dd}u|^2}{1+|u|^2}+b \thickbar{D}c-\thickbarsmall{c}D \thickbarsmall{b}+{\vkappa \over 2(1+|u|^2)}\,b  c \thickbarsmall{b}\thickbarsmall{c}\,,
\eea
with Christoffel symbols $\Gamma^u_{uu}=-\frac{\thickbarsmall{u}}{1+|u|^2}$. 
This is the so-called SUSY cigar model: its bosonic counterpart appeared in~\cite{WittenCigar}, and the SUSY model was studied, for instance, in~\cite{JackJonesCigar, Hori:2001ax, KarchTong}. We will leave a detailed study of this model from the perspective of the GN formulation for the future.

\begin{figure}
\centering
\begin{overpic}[abs, scale=1,unit=1mm]{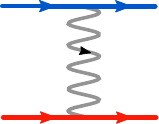}
    \put(-3,21){$p$}
    \put(28,21){$p$}
    \put(29,0){$q=0$}
    \end{overpic}
    \vspace{0.2cm}
\caption[4ptfunctree]{Tree level contribution to the 4-point function of bosonic fields with special kinematics: the momentum flowing through the lower line is taken to be zero.}
\label{fig4ptfunctree}
\end{figure}

\section{Beta function}\label{betafuncsec} 

\nd The goal of this section is to compute the $ \beta $-function of the proposed superdeformed model~(\ref{Def_Lagrangian}) at one and two loops. The corresponding Feynman rules are shown in Fig.~\ref{feynfig}.

In the course of this we shall address the constraints on the quartic interaction in \eqref{Def_Lagrangian} that ensure renormalizability of the model. In fact, we will do it in greater generality than strictly needed for the model~(\ref{Def_Lagrangian}), by allowing the current $J$ to take values in an arbitrary complex Lie algebra $\mathfrak{g}_{\CC}$ and assuming that $r_{\fk{s}}\in \mathrm{End}(\mathfrak{g}_{\CC})$ is an arbitrary linear operator (varying smoothly with $\fk{s}$). We introduce a basis of Hermitian unit-normalized generators $\tau_a$ in this Lie algebra ($\mathrm{Tr}(\tau_a\tau_b)=\delta_{ab}$) and define the structure constants via $[\tau_a, \tau_b]=i\,f_{ab}^c\,\tau_c$. Ultimately we shall set $\mathfrak{g}_{\CC}=\mathfrak{sl}_2$, in which case $\tau_a={\sigma_a\over \sqrt{2}}$ and $f_{ab}^c=\sqrt{2}\epsilon_{abc}$, where $\sigma_a$ are the Pauli matrices.

In the present section we will be extracting the $\beta$-function from the 4-point correlation function with fixed external momenta, where, moreover, the momentum through the lower line is taken to be zero (for more on this see~\cite{BykovBeta}). The tree level expression for the connected 4-point function, as shown in Fig.~\ref{fig4ptfunctree}, is
	\bea\label{G4tree}
	G_4^{\mathrm{tree}}=-\vkappa\,\sum\limits_{a}\,r_{\fk{s}}(\tau_a)\otimes \tau_a\,.
	\eea
We now proceed to compute one- and two-loop corrections.
\enlargethispage{1cm}


	\subsection{One loop} 
	
	\begin{figure}
\centering
\begin{overpic}[abs, scale=1,unit=1mm]{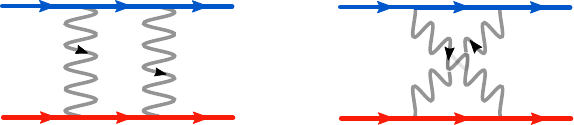}
    \end{overpic}
    \vspace{0.2cm}
\caption[Fig1loopbeta]{One-loop contributions to the 4-point function.}
\label{fig1loopbeta}
\end{figure}
	At one loop one has the two diagrams shown in Fig.~\ref{fig1loopbeta}. Summing these two diagrams, one gets the UV divergent result\footnote{Here $(p, z)\equiv pz+\thickbarsmall{p}\,\thickbarsmall{z}$ is the scalar product in 2D. Although we use the same letter for a two-vector and for its holomorphic component, the scalar product with a two-vector is always denoted by round brackets.}
	\bea\label{G4oneloop}
	G_4^{\mathrm{1\tx{-}loop}}=
	-{\vkappa^2} \int \,\frac{d^2 z}{(2\pi)^2}\frac{e^{i(p, z)}}{|z|^2}\times {1\over 2}\sum\limits_{a, b}\, [r_{\fk{s}}(\tau_a), r_{\fk{s}}(\tau_b)]  \otimes [\tau_b, \tau_a]
	\eea
	In~\cite{BykovBeta} we showed that 
	\bea
	\mathsf{A}(p):={1\over 2\pi}\int \,d^2 z\frac{e^{i(p, z)}}{|z|^2}=-{1\over 2}\log{\left({p^2 \varepsilon^2}\right)}+\textrm{finite terms}\,,
	\eea
	 where $\varepsilon$ is the UV cutoff. 
	Simplifying the last term in~(\ref{G4oneloop}), 	$
	[r_{\fk{s}}(\tau_a), r_{\fk{s}}(\tau_b)]\otimes [\tau_b, \tau_a]=-i\,f_{ab}^c\,[r_{\fk{s}}(\tau_a), r_{\fk{s}}(\tau_b)]\otimes \tau_c\,, 
$
	we derive the condition for the renormalizability of the model at one loop:
	\bea\label{renormeq}
	\boxed{
	\quad -{\vkappa^2\over 4\pi}\,i\,f_{ab}^c\,[r_{\fk{s}}(\tau_a), r_{\fk{s}}(\tau_b)]=\vkappa\,\upbeta_{\fk{s}}^{\mathrm{1\tx{-}loop}}(\fk{s}, \vkappa)\, \dot{r}_{\fk{s}}(\tau_c)+\upbeta_{\vkappa}^{\mathrm{1\tx{-}loop}}(\fk{s}, \vkappa) \,r_{\fk{s}}(\tau_c)\,,\quad 
	}
	\eea
where $\dot{r}_{\fk{s}}:=\fk{s} {d r_{\fk{s}}\over d \fk{s}}$, and 
the $\beta$-functions of the two couplings are defined by\footnote{Recall that we have chosen the RG time direction $t\in (-\infty, 0)$  pointing towards the IR, so that $t=\log{\varepsilon}$ with $\varepsilon$ the coordinate space cutoff.}
	\bea 
	\upbeta_{\fk{s}}={d\log{\fk{s}}\over d \log{\varepsilon}}\,,\quad\quad \upbeta_{\vkappa}={d\vkappa\over d \log{\varepsilon}}\,.
	\eea
	The loop expansion here and below refers to the expansion in $\vkappa$, so that at one loop $\upbeta_{\fk{s}}^{\mathrm{1\tx{-}loop}}(\fk{s}, \vkappa)=\mathcal{O}(\vkappa)$ and $\upbeta_{\vkappa}^{\mathrm{1\tx{-}loop}}(\fk{s}, \vkappa)=\mathcal{O}(\vkappa^2)$. As a result, $\vkappa$ dependence drops out of~(\ref{renormeq}), hence it reduces to a differential equation in $\fk{s}$. 
	
	\vspace{0.4cm}
	\textbf{The $\mathfrak{sl}_2$ case.} Expression~(\ref{renormeq}) is the constraint for one-loop renormalizability of the model valid for an arbitrary Lie algebra. In order to see whether our $\mathfrak{sl}_2$ $r$-matrix satisfies this relation we recall the remarkable property\footnote{One way of deriving this relation is by considering the $u\to 1$ limit of the classical Yang-Baxter equation
	\begin{equation*}\label{CYBE}
		\left[ r_{\fk{s}}[A] , r_{\fk{s} u}[B] \right] + r_{\fk{s}}\left[ \left[ r_{u}[B], A \right] \right] + r_{\fk{s} u} \left[ \left[ B, r_{u^{-1}}[A] \right] \right] = 0 \, ,
	\end{equation*}
	taking into account that the $r$-matrix has the asymptotic form 
	$r_u={1\over 2}{1+u\over 1-u} \,\mathrm{Id}+\mathcal{O}(u-1)$
	 at the singular point $u=1$. This is only true in the $\mathfrak{sl}_2$ case~(\ref{rmatrixdef}) but not for higher $n$.} 
	 (\ref{rmatalgsu2_0}) that holds for the $\mathfrak{sl}_2$ $r$-matrix as defined in~(\ref{rmatrixdef}):
	\bea\label{rmatalgsu2}
	\dot{r}_{\fk{s}}([\mathbb{A}, \mathbb{B}])=[r_{\fk{s}}(\mathbb{A}), r_{\fk{s}}(\mathbb{B})]\,.
	\eea
	Setting $\mathbb{A}=\tau_a$ and $\mathbb{B}=\tau_b$, one gets from~(\ref{rmatalgsu2}) $i\,\sqrt{2}\epsilon_{abc}\,\dot{r}_{\fk{s}}(\tau_c)=[r_{\fk{s}}(\tau_a), r_{\fk{s}}(\tau_b)]$, which ensures that~(\ref{renormeq}) holds with the following $\beta$-functions:
	\bea\label{1loopbetafunc}
	\upbeta_{\fk{s}}^{\mathrm{1\tx{-}loop}}={\vkappa\over \pi}\,,\quad\quad \upbeta_{\vkappa}^{\mathrm{1\tx{-}loop}}=0\,.
	\eea 
		
	Thus, (\ref{rmatalgsu2}) underpins the renormalizability of the deformed model. Summarizing, for~(\ref{G4oneloop}) we get the expression
	\bea\label{1loopfin}
	G_4^{\mathrm{1\tx{-}loop}}=
	-{\vkappa^2\over \pi} \mathsf{A}(p)\;\sum\limits_{a}\, \dot{r}_{\fk{s}}(\tau_a)\otimes \tau_a\,.
	\eea

\vspace{0.3cm}
\textbf{Remark.} Curiously, in the $\mathfrak{sl}_2$-case, and for $\upbeta_{\fk{s}}^{\mathrm{1\tx{-}loop}}=\mathrm{const.}, \,\upbeta_{\vkappa}^{\mathrm{1\tx{-}loop}}=0$, equations~(\ref{renormeq}) are equivalent to Nahm equations~\cite{Nahm} from the theory of monopoles. Asymptotically conformal theories correspond to solutions defined on a semi-infinite line: $t\in (-\infty, 0)$. These have been studied, from a mathematical perspective, in~\cite{Kronheimer} and, from a more physical one, in~\cite{GaiottoWitten}. It appears that in the $\mathfrak{sl}_2$-case the solution is unique and corresponds to the $r$-matrix studied here (in the context of monopoles it was found in~\cite{Nahm}).

\enlargethispage{0.5cm}


\subsection{Two loops}

\begin{figure}[h]
	\centering
	\begin{overpic}[abs, scale=0.8,unit=1mm]{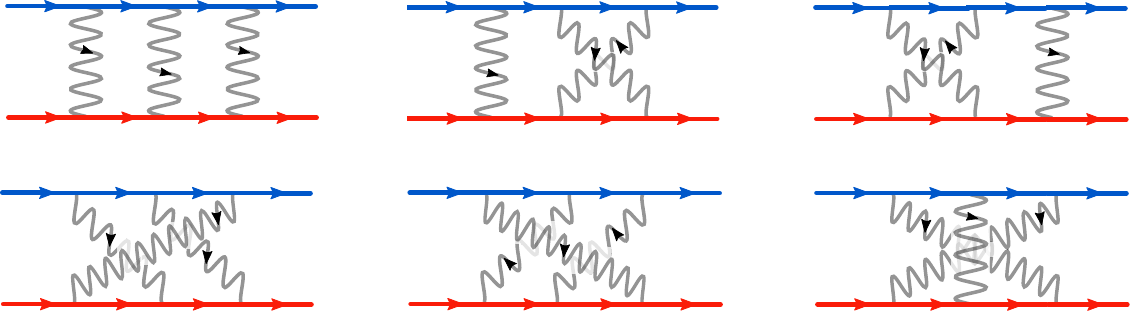}
	\end{overpic}
	\vspace{0.2cm}
	\caption[Fig2loopbeta]{Two-loop contributions to the beta function.}
	\label{fig2loopbeta}
\end{figure}

At two loops one has $ 3! $ diagrams as indicated in Fig.~\ref{fig2loopbeta}. The 2-loop integrand is\footnote{Again, we start with an arbitrary Lie algebra and specialize to $\mathfrak{sl}_2$ later on.} 
\begin{equation}\label{key}
	\begin{aligned}
		\cl{I}_{4}^{\mathrm{2\tx{-}loop}} & =  -\vkappa^{3}\, e^{i\left(p,z_{13}\right)} \times \frac{r_{\fk{s}}[\tau_a] r_{\fk{s}}[\tau_b] r_{\fk{s}}[\tau_c]}{z_{12} z_{23}} \\ 
		& \otimes \left(\frac{\tau_a \tau_b \tau_c}{\thickbarsmall{z}_{12} \thickbarsmall{z}_{23}} + \frac{\tau_a \tau_c \tau_b}{\thickbarsmall{z}_{13} \thickbarsmall{z}_{32}} + \frac{\tau_b \tau_a \tau_c}{\thickbarsmall{z}_{21} \thickbarsmall{z}_{13}} + \frac{\tau_c \tau_a \tau_b}{\thickbarsmall{z}_{31} \thickbarsmall{z}_{12}} + \frac{\tau_b \tau_c \tau_a}{\thickbarsmall{z}_{23} \thickbarsmall{z}_{31}} + \frac{\tau_c \tau_b \tau_a}{\thickbarsmall{z}_{32} \thickbarsmall{z}_{21}}\right) \,,
	\end{aligned}
\end{equation} 
which upon simplifying 
produces the following:  
\begin{equation}\label{G4twoloop}
	\begin{aligned}
		&&G_4^{\mathrm{2\tx{-}loop}}=
		-\vkappa^3\, \int \,\frac{d^2 z_{12}}{(2\pi)^2}\,\frac{d^2 z_{23}}{(2\pi)^2}\frac{e^{i(p, z_{13})}}{z_{12}z_{23}}\,\sum\limits_{a, b, c}\,r_{\fk{s}}(\tau_a)r_{\fk{s}}(\tau_b)r_{\fk{s}}(\tau_c) \\ &&\hspace{2cm}\otimes\left({1\over \thickbarsmall{z}_{12}\thickbarsmall{z}_{23}}\,[\tau_a, [\tau_b, \tau_c]]-{1\over \thickbarsmall{z}_{12}\thickbarsmall{z}_{13}}\,[\tau_b, [\tau_a, \tau_c]]\right) \, .
	\end{aligned}
\end{equation}
First of all, let us compute the divergent parts of these integrals. The integral of the first term in~(\ref{G4twoloop}) is proportional to $\mathsf{A}(p)^2$\,, so the only nontrivial computation has to do with the double integral
\begin{equation}
\mathsf{B}(p)={1\over (2\pi)^2}\int \,d^2 z_{12}\,d^2 z_{23}\,\frac{e^{i (p, z_{13})}}{|z_{12}|^2 z_{23} \thickbarsmall{z}_{13}}
\end{equation}
Let us assume that we regulate both integrals by imposing UV cutoffs ${|z_{12}|>\varepsilon},$ ${|z_{23}|>\varepsilon}$. Next, we compute the derivative 
\begin{equation}
	\begin{aligned}
		{\dd\mathsf{B}(p) \over \dd \thickbarsmall{p}}= \underbracket[0.65pt][1.5pt]{\int\limits_{|z_{12}|>\epsilon} \,{d^2 z_{12}\over 2\pi}\,\frac{e^{i (p, z_{12})}}{|z_{12}|^2 }}_{\mathsf{A}} \,\, \times \,\, \underbracket[0.65pt][1.5pt]{i\!\!\!\!\int\limits_{|z_{23}|>\varepsilon} \,{d^2 z_{23}\over 2\pi}\,\frac{e^{i (p, z_{23})}}{z_{23}}}_{\frac{\partial\mathsf{A}}{\partial\thickbarsmall{p}}} = \dfrac{1}{2} \dfrac{\partial}{\partial \thickbarsmall{p}} \, \mathsf{A}(p)^{2} 
	\end{aligned}
\end{equation}
where we have used the identity $z_{13}=z_{12}+z_{23}$, hence
\bea
\mathsf{B}(p)={1\over 2} \mathsf{A}(p)^2+\textrm{finite terms}
\eea
Thus, up to finite terms, one can write the two-loop contribution~(\ref{G4twoloop}) as follows:
\bea\label{2loopdiv0}
G_4^{\mathrm{2\tx{-}loop}}\simeq-{\vkappa^3\over (2\pi)^2}\, \mathsf{A}(p)^2\,\sum\limits_{a, b, c}\,r_{\fk{s}}(\tau_a)r_{\fk{s}}(\tau_b)r_{\fk{s}}(\tau_c) \otimes\underbracket[0.65pt][1.5pt]{\left(\,[\tau_a, [\tau_b, \tau_c]]-{1\over 2}\,[\tau_b, [\tau_a, \tau_c]]\right)}_{={1\over 2}\left(\,[\tau_a, [\tau_b, \tau_c]]+[[\tau_a, \tau_b], \tau_c]\right)} \, ,
\eea
where we have used the Jacobi identity to simplify the term in brackets. 

It follows from~(\ref{2loopdiv0}) that the two loop divergence is proportional to the square of the one loop divergence~(\ref{1loopfin}). Using the general expression~(\ref{renormeq}), one can then show that it may be removed by the renormalization of the coupling constants at one loop. For simplicity, though, from now on we restrict to the case $\mathfrak{g}_{\CC}=\mathfrak{sl}_2$, where the proof is considerably more succinct. Recalling~(\ref{rmatalgsu2}), which can be rewritten as ${\dot{r}_{\fk{s}}(\tau_a)=-{i\over 2\sqrt{2}}\epsilon_{abc}\,[r_{\fk{s}}(\tau_b), r_{\fk{s}}(\tau_c)]}$\,, we recast~(\ref{2loopdiv0}) as 
\bear\nonumber
&&G_4^{\mathrm{2\tx{-}loop}}={\vkappa^3\over (2\pi)^2}\, \mathsf{A}(p)^2\,\sum\limits_{a, b}\,\left(r_{\fk{s}}(\tau_a)\dot{r}_{\fk{s}}(\tau_b)+\dot{r}_{\fk{s}}(\tau_a)r_{\fk{s}}(\tau_b)\right) \otimes [\tau_a, \tau_b]=\\ \label{2loopdiv}&&=-{\vkappa^3\over 2\pi^2}\, \mathsf{A}(p)^2\,\,\sum\limits_{a}\,\ddot{r}_{\fk{s}}(\tau_a) \otimes \tau_a
\eear
This result implies renormalizability at two loops and shows that there is no two-loop contribution to the $\beta$-function\footnote{The fact that, at two loops, the sum of ladder diagrams is proportional to the \emph{square} of the sum of one-loop diagrams was first observed (in the undeformed case) in~\cite{DestriBeta}.}. Clearly, the divergent term in~(\ref{2loopdiv}) is eliminated by the same shift of parameter $\fk{s}$ in the tree-level expression~(\ref{G4tree}) that cancels the one-loop divergence in~(\ref{1loopfin}). Thus, $\beta_{\fk{s}}^{\mathrm{2\tx{-}loop}}=\beta_{\vkappa}^{\mathrm{2\tx{-}loop}}=0$. 

To summarize, the $\beta$-functions~(\ref{1loopbetafunc}) are exact up to two loops and lead to the RG flow equations~(\ref{RGflow}), whose solution is schematically depicted in Fig.~\ref{RGFT}. The flow interpolates between the $\fk{s}=0$ limit in the UV, which will be discussed in detail in the next section, and the round metric on $\CP^1$ in the IR limit $\fk{s}=1$.

\vspace{0.3cm}
\textbf{Remark.} This result is in agreement with \cite{Gamayun} (as well as with earlier work~\cite{LeClair}), where $ \beta $-functions  for generic 2D sigma models were investigated in the first order formalism. An important special case considered therein is that of ``Lie-algebraic sigma models'', which are defined by the requirement that the inverse target space metric may be written as $G^{a\thickbarsmall{a}}=G^{A\thickbarsmall{A}} v^a_A v^{\thickbarsmall{a}}_{\thickbarsmall{A}}$ where $v^a_A(U) {\dd \over \dd U^a}$ are vector fields generating a Lie algebra $\mathfrak{g}_{\CC}$ of some complex Lie group $G_{\CC}$ acting on the target space (accordingly, the index $A$ takes the values $A=1, \ldots, \mathrm{dim} \, \mathfrak{g}_{\CC}$). Models with such metrics were introduced in~\cite{CY}. The Gross-Neveu models considered in the present paper are special cases where the vector fields $v_a^A$ are linear in $U^a$. In this case the Hamiltonian (the interaction term in~(\ref{Def_Lagrangian})) may be written as $G^{a\thickbarsmall{a}}V_a V_{\thickbarsmall{a}}=G^{A\thickbarsmall{A}}J_AJ_{\thickbarsmall{A}}$, where $J=J_A \tau_A$ is a $\mathfrak{g}_{\CC}$-valued Kac-Moody current with the OPE
\begin{equation}\label{currOPE}
	J_{A}(z)J_{B}(w) = \dfrac{\eta_{AB}}{(z-w)^{2}} + i\dfrac{f_{AB}^{C}J_{C}}{z-w} + \;\cdots 
\end{equation} 
For this class of models the following $\beta^{\tx{2-loop}}$ function was found in\footnote{It has been shown in~\cite{Gamayun2} that, in the case when the vector fields $v_a^A$ are \emph{not} linear, the definition of $\eta_{AB}$ needs to be amended.}~\cite{Gamayun}:
\bear
\label{Beta2L_GN}
	&&\beta_{\mathrm{2\tx{-}loop}}^{C\thickbarsmall{C}} = \dfrac{1}{2} G^{A\thickbarsmall{A}} G^{B\thickbarsmall{B}} G^{C\thickbarsmall{C}} \left( f_{AE}^{D} f_{BD}^{E} \eta_{\thickbarsmall{A}\thickbarsmall{B}} + \textrm{c.c.} \right)\,.
\eear

In case of a simple Lie algebra one may take $\eta_{AB}=k\,\delta_{AB}$, where $k$ is the level of the current algebra. In the case of the supersymmetric models the level is zero, as can be inferred from the current \eqref{currdef}, so that $ \eta_{AB} =\eta_{\thickbarsmall{A}\thickbarsmall{B}}= 0 $, implying the vanishing of the two-loop contribution~(\ref{Beta2L_GN}).

	\section{Super-Thirring model in superspace}\label{Super_Thirring_Superspace}
	
	In the present section we will consider the models resulting in the UV limit, $\fk{s}\to 0$, of the deformed SUSY $\CP^1$ sigma model. These are two conformal field theories on the two sides of the correspondence:
	\begin{itemize}
	\item The super-Thirring model \cite{Freedman1, Freedman2} that arises in the $\fk{s}\to 0$ limit of the Gross-Neveu model~(\ref{GNinhom}). It is  defined by the Lagrangian 
	\begin{equation}\label{SUSYThirring}
	\mathcal{L}=v\,\thickbar{\dd}u+\thickbarsmall{u}\,\dd \thickbarsmall{v}+b\,\thickbar{\dd}c-\thickbarsmall{c}\,\dd \thickbarsmall{b}+{\vkappa\over 2} \,\big|v u+b c\big|^2
	\end{equation}
	\item The sigma model with target space the (super)-cylinder, with bosonic part $\mathbb{R}\times S^1$, that arises on the sigma model side and is described by the free Lagrangian~(\ref{cylfreetheory}).
	\end{itemize}
Conjecturally this is a complete equivalence of CFTs. In support of this we will outline the computation of correlation functions of the elementary fields for both CFTs in the general case and explicitly compute the 2- and 4-point functions.

\vspace{0.20cm}
\textbf{Remark.} A complete characterization of the CFTs would involve the construction of the energy-momentum tensor.
Here a natural puzzle arises. On the one hand, the theory~(\ref{cylfreetheory}) is free, so its central charge is independent of $\vkappa$. On the other,  (\ref{SUSYThirring}) is an interacting theory, whose central charge was computed in~\cite{Lauer} (cf. also~\cite{Freedman1} for the relevant methods) and is linear in $\vkappa$: $c=c_0+c_1 \vkappa$, where $c_{0}$ and $c_{1}\neq 0$ are some numerical constants. We believe that the $\vkappa$-dependent mismatch is cancelled by a linear dilaton profile along the cylinder, i.e. $\Phi\sim W+\thickbar{W}$, which generates an additional term $\delta T_{zz}\sim \dd^2 \left(W+\thickbar{W}\right)$ in the energy-momentum tensor. Since the correlator $ \langle W \thickbar{W}\rangle= \cl{O}\left[ \vkappa \right] $, it leads to a contribution to the central charge that is linear in $\vkappa$. The origin of this linear dilaton term is presumably the Jacobian coming from the integration over $v, \thickbarsmall{v}$ that one performs in order to pass from~(\ref{SUSYThirring}) to the cylinder sigma model~(\ref{cylfreetheory}).

\subsection{On-shell supersymmetry}

	First of all, let us establish that the model~(\ref{SUSYThirring}) is supersymmetric. We will follow the same strategy as in the non-Abelian case in section~\ref{Susy_Def_Sec}. It is immediate to see that free part of the Lagrangian above is invariant (up to a total derivative) w.r.t. the transformations~(\ref{Supersymmetry_1}):
	\bea\label{SUSYtrans}
	\delta u=\epsilon_1 c\,,\quad\quad \delta b=- \epsilon_1 v\,,\quad\quad\delta c=-\epsilon_2 \dd u\,,\quad\quad \delta v=\epsilon_2 \dd b\,,
	\eea
	and similarly for the conjugate variables. These transformations are known since the seminal work~\cite{FMS} (and have reappeared in the work on abstract $\beta\gamma$-systems in~\cite{Kapustin, Policastro}), and supersymmetry of the interacting model~(\ref{SUSYThirring}) was established in~\cite{Freedman1, Freedman2}. 
	
	To show invariance of the interaction term, we introduce the notation
	\bea
	J=v u+b c\,,\quad\quad \hat{J}=bu\,.
	\eea
	Here $J$ is the $\CC^\ast$ Kac-Moody current of the system, and $\hat{J}$ its superpartner, since $\delta J=\epsilon_2 \dd \hat{J}$.
	
	The interaction term in~(\ref{SUSYThirring}) can be written as~${\vkappa\over 2} J \thickbarsmall{J}$. 
	 The peculiarity of this system is that $J$ is holomorphic both in the free and interacting systems:
	\bea\label{currcons}
	\thickbar{\dd} J=\dd \thickbarsmall{J}=0\,.
	\eea
The variation of the interaction part of the action w.r.t. SUSY transformations  is
	\bea\label{actionSUSYvar}
	\delta \int\,d^2z\, J \thickbarsmall{J}=\int\,d^2z\, \epsilon_2 \dd \hat{J} \thickbarsmall{J}\,,
	\eea
	which is easily seen to be zero on-shell upon integrating by parts and using~(\ref{currcons}).
	
	It is perhaps more convincing when we make SUSY transformations off-shell\footnote{Here we follow the same route as in~\cite{BykovSUSY}.}. To this end we notice that~(\ref{actionSUSYvar}) looks like the variation of the action under a $\CC^\ast$ gauge transformation of the conjugate fields with gauge parameter $\epsilon_2 \hat{J}$ (only the free part of the action is non-invariant). It can therefore be compensated by an opposite gauge transformation as follows:
	\bea\label{compenstrans}
	\hat{\delta} \thickbarsmall{u}= {\vkappa\over 2}\,\epsilon_2 \hat{J}\thickbarsmall{u}\,,\quad\quad   \hat{\delta} \thickbarsmall{v}= -{\vkappa\over 2}\,\epsilon_2 \hat{J}\thickbarsmall{v}\,,\quad\quad \hat{\delta} \thickbarsmall{c}= {\vkappa\over 2}\,\epsilon_2 \hat{J}\thickbarsmall{c}\,,\quad\quad \hat{\delta} \thickbarsmall{b}= -{\vkappa\over 2}\,\epsilon_2 \hat{J}\thickbarsmall{b}
	\eea
	The complete off-shell SUSY transformation is obtained by combining~(\ref{SUSYtrans}) and~(\ref{compenstrans}):
	\bea
	\delta_{\mathrm{os}}=\delta+\hat{\delta}
	\eea
	Note however that the transformations so defined only satisfy the SUSY algebra on-shell. For example, using $\delta \hat{J}=-\epsilon_1 J$ and $\hat{J}^2=0$, we get $\delta(\delta  \thickbarsmall{u})\sim {\vkappa\over 2} J \thickbarsmall{u}=\dd \thickbarsmall{u}$, where in the last step we have used the e.o.m. As usual, to close the algebra off-shell one should introduce auxiliary fields, which is done the easiest by passing to superspace.
	
	
\subsection{$\mathcal{N}=(1, 1)$ superspace} We start with $\mathcal{N}=(1, 1)$ superspace\footnote{This nomenclature is borrowed from Minkowski space. In the present Euclidean setup there is a complex left-moving supercharge and its complex-conjugate right-moving one. For details on Euclidean superspace cf.~\cite[Chapters 22, 23]{West}.}: here one has two super-coordinates $\theta, \thickbarsmall{\theta}$ and the following supercharges/superderivatives:
\bear\label{11supercharge}
Q={\dd\over \dd \theta}+\theta {\dd\over \dd z}\,,\quad\quad \thickbarsmall{Q}={\dd\over \dd \thickbarsmall{\theta}}+\thickbarsmall{\theta} {\dd\over \dd \thickbarsmall{z}}\,,\quad\quad
D={\dd\over \dd \theta}-\theta {\dd\over \dd z}\,,\quad\quad \thickbarsmall{D}={\dd\over \dd \thickbarsmall{\theta}}-\thickbarsmall{\theta} {\dd\over \dd \thickbarsmall{z}}
\eear
Let us also introduce two generic superfields
	\bea\label{UBdecomp}
	\bm{U}:=U_0+\thickbarsmall{\theta} \,U_1\,,\quad\quad \bm{B}:=B_0+\thickbarsmall{\theta} \,B_1\,,
	\eea
	where $U_{0, 1}$ and $B_{0, 1}$ are functions of $\theta$. Postulating that $\bm{U}$ be a commuting field, whereas $\bm{B}$ be anti-commuting, we write the holomorphic part of the free Lagrangian as follows\footnote{Our convention for integration is that $\int d\theta d\thickbarsmall{\theta}\,\bullet \equiv\int d\theta \left(\int d\thickbarsmall{\theta}\,\bullet \right)$, with an obvious extension in the case of more superspace coordinates.} (here $d^2\theta\equiv d\theta d\thickbarsmall{\theta}$):
	\bea
	\mathcal{L}=-\int\,d^2\theta\, \bm{B}  \thickbarsmall{D}  \bm{U} =-\int\,d\theta\,\left(B_0 \thickbarsmall{\dd} U_0+B_1 U_1 \right)
	\eea
	It follows that $B_1, U_1$ are auxiliary fields and may be safely dropped. Decomposing the residual fields as 
	\bea
	U_0=u+\theta c\,,\quad\quad B_0=b-\theta v\,
	\eea
	and integrating over $\theta$, one finds the Lagrangian of the $\beta\gamma$-system 
	\bea
	\mathcal{L}=v \,\thickbarsmall{\dd} u+b \,\thickbarsmall{\dd} c\,.
	\eea
	It is also clear that the supercharge $Q$ from~(\ref{11supercharge}), when acting on $U_0, B_0$, reproduces the transformation laws~(\ref{SUSYtrans}) for $\epsilon_2=\epsilon_1$. The supercharge $\thickbarsmall{Q}$ shifts the fields $U_0, B_0$ by auxiliary fields that are zero on-shell.

	What remains is to describe interactions. The claim is that the full Lagrangian takes the form 
	
	\bea\label{11SUSYlagr}
	\mathcal{L}=-\int\,d^2\theta\, \left(\bm{B}  \thickbarsmall{D}  \bm{U}+\thickbarsmall{\bm{U}}D \thickbarsmall{\bm{B}}+{\vkappa\over 2}\,  \bm{J} \thickbarsmall{\bm{J}}\right)\,,
	\eea
	where $\bm{J}=\bm{U} \bm{B}$ 
	is the current superfield. In the interacting case auxiliary fields are non-zero and are crucial to get correct SUSY transformations. Indeed, one has, for  example,
	\bea
	\delta \bm{U}=(\epsilon Q+\thickbarsmall{\epsilon} \thickbarsmall{Q})\bm{U}=\epsilon Q U_0+\thickbarsmall{\theta} \epsilon Q U_1+\thickbarsmall{\epsilon} \, U_1-\thickbarsmall{\theta} \,\thickbarsmall{\epsilon} \,\thickbarsmall{\dd} U_0
	\eea
	implying that the transformation law of the dynamical field $U_0$ is
	\bea\label{N11SUSYvar}
	\delta U_0=\epsilon Q U_0+\thickbarsmall{\epsilon} U_1\,.
	\eea
	As already mentioned, the first term is the simple transformation law~(\ref{SUSYtrans}) (for $\epsilon_2=\epsilon_1=\epsilon$), whereas the second piece depends on the auxiliary field $U_1$. We will show in Appendix~\ref{auxfieldsapp} that, upon elimination of the auxiliary fields, one arrives precisely at~(\ref{compenstrans}) and its complex conjugate.

\subsection{$\mathcal{N}=(2, 2)$ superspace}

In this subsection we develop $(2,2)$-superspace framework and begin by introduction of the corresponding supercharges and superderivatives
	\bear
	&&Q_1={\dd\over \dd \theta_1}+{1\over 2}\theta_2 {\dd\over \dd z}\,,\quad\quad Q_2={\dd\over \dd \theta_2}+{1\over 2}\theta_1 {\dd\over \dd z}\,,\;\\
	&&D_1={\dd\over \dd \theta_1}-{1\over 2}\theta_2 {\dd\over \dd z}\,,\quad\quad D_2={\dd\over \dd \theta_2}-{1\over 2}\theta_1 {\dd\over \dd z}\,,
	\eear
	together with their complex conjugates. 

	Clearly, 
	\bea
	Q_1^{\,2}=Q_2^{\,2}=0=\thickbarsmall{Q}_1^{\,2}=\thickbarsmall{Q}_2^{\,2}\,\quad\quad \textrm{and}\quad\quad  \{Q_1, Q_2\}= {\dd\over \dd z}\,,\quad\quad  \{\thickbarsmall{Q}_1, \thickbarsmall{Q}_2\}= {\dd\over \dd \thickbarsmall{z}}
	\eea
	so that indeed one has the $(2, 2)$ SUSY algebra.

We will define two complex superfields $\bm{U}, \bm{B}$ as before, one commuting and the other anti-commuting, now the difference being that one is chiral and the other twisted chiral, i.e.\footnote{As a result of these definitions, $\bm{B}$ and $\thickbarsmall{\bm{B}}$ are both twisted chiral. It has been observed in the literature~\cite{BuscherRocek, West, HullLindstrom} that this is an obstacle for K\"ahler potential type terms like $\int d^4\theta \bm{B} \thickbarsmall{\bm{B}}$ in the Lagrangian, since these are automatically total derivatives. Nevertheless, certain interesting Lagrangians like~(\ref{UB3thetaint}) below may be constructed using such fields.}
\bear
&&D_2\bm{U}=0\,,\quad \thickbarsmall{D}_1\bm{U}=0\,\quad\quad \textrm{(chiral)}\\
&&D_2\bm{B}=0\,,\quad\thickbarsmall{D}_2\bm{B}=0\,\quad\quad \textrm{(twisted chiral)}
\eear
Upon introducing the combinations $z_-=z-{1\over 2}\theta_1\theta_2$ and $\thickbarsmall{z}_{\pm}=\thickbarsmall{z}\pm {1\over 2}\thickbarsmall{\theta}_1 \thickbarsmall{\theta}_2$, the constraints are solved as follows in terms of component fields:
\bear
&&\bm{U}=u(z_-, \thickbarsmall{z}_+)+\theta_1 \,c(z_-, \thickbarsmall{z}_+)+\thickbarsmall{\theta}_2\,d(z_-, \thickbarsmall{z}_+)+\theta_1\thickbarsmall{\theta}_2\,f(z_-, \thickbarsmall{z}_+)\,,\\
&&\bm{B}=b(z_-, \thickbarsmall{z}_-)-\theta_1\,v(z_-, \thickbarsmall{z}_-)+\thickbarsmall{\theta}_1\,w(z_-, \thickbarsmall{z}_-)+\theta_1 \thickbarsmall{\theta}_1\,g(z_-, \thickbarsmall{z}_-)\,.
\eear
To convince oneself that this is the correct superfield content for our model, one can work out the transformation properties of the fields under, say, left-moving supersymmetry (transformation properties w.r.t. $\thickbarsmall{Q}_1, \thickbarsmall{Q}_2$ may be computed in a similar way): 
\bear
&&(\epsilon_1 Q_1+\epsilon_2 Q_2) \bm{U}=\epsilon_1 c-\thickbarsmall{\theta}_2 \epsilon_1 f-\theta_1 \epsilon_2 \dd u + \theta_1\thickbarsmall{\theta}_2\,\epsilon_2\dd d \,,\\
&&(\epsilon_1 Q_1+\epsilon_2 Q_2) \bm{B}=-\epsilon_1 v-\thickbarsmall{\theta}_1 \epsilon_1 g-\theta_1 \epsilon_2 \dd b+ \theta_1\thickbarsmall{\theta}_1\,\epsilon_2\dd w
\eear
so that, in particular,
\bear
\delta u=\epsilon_1 c\,,\quad\quad \delta c=- \epsilon_2 \dd u\,,\quad\quad
\delta b=-\epsilon_1 v\,,\quad\quad \delta v= \epsilon_2 \dd b\,,
\eear
which correctly reproduces the transformation law~(\ref{SUSYtrans}).  

The product $\bm{U}\bm{B}$ is a (left) semi-chiral field\footnote{Twisted chiral and semi-chiral fields were introduced in~\cite{GatesRocek} and~\cite{BuscherRocek} respectively.}, i.e. $D_2\left(\bm{U}\bm{B}\right)=0$, so that one can form the density
\bea\label{UB3thetaint}
\mathcal{L}_0=\int\,d^3\theta\;\bm{U}\bm{B}={1\over 2}\left(-u \, \thickbarsmall{\dd}v+v \,\thickbarsmall{\dd}u+c \,\thickbarsmall{\dd}b + b \, \thickbarsmall{\dd} c \right)+f \,w-d \, g\,,
\eea
where $d^3\theta=d\theta_1 d\thickbarsmall{\theta}_1 d\thickbarsmall{\theta}_2$. Thus, up to the last two terms involving auxiliary fields, one obtains the free holomorphic part of the action.

To incorporate the interaction term, we will use the SUSY St\"uckelberg formalism. Recall that the usual Thirring model may be written as $\mathcal{L}=i \, \thickbarsmall{\Psi} \slashed{D}\Psi+A_{\mu}^2$, which upon integration over the auxiliary fields $A_\mu$ generates the four-fermion interaction. This can be formally promoted to a gauge-invariant theory by introducing an extra scalar St\"uckelberg field $s$ and replacing the Lagrangian with $\mathcal{L}=i \, \thickbarsmall{\Psi} \slashed{D}\Psi+\left(A_{\mu}-\partial_\mu s\right)^2$. Since gauge transformations act on  $s$ by simple shifts, one can pick the gauge $s=0$, thus returning to the original model. In the SUSY setup the gauge-invariant extension is useful, since it allows passing to an analogue of Wess-Zumino gauge.

To implement this idea, observe that the free action~(\ref{UB3thetaint}) has a global symmetry
\bea
\bm{U} \to \uplambda \bm{U}, \quad\quad\bm{B}\to \uplambda^{-1} \bm{B}\,,\quad\quad \textrm{where} \quad\quad\uplambda \in \CC^\ast\,. 
\eea
In fact, this is still a symmetry if one assumes that $\uplambda=\uplambda(z)$, which corresponds to the Kac-Moody invariance of~(\ref{UB3thetaint}). To gauge this symmetry, we introduce a semi-chiral gauge field\footnote{It was observed in~\cite{RocekLinear} that one can gauge Kac-Moody symmetries using semi-chiral superfields.} $\bm{V}$: 
\bea
\mathcal{L}_{\mathrm{gauged}}=\int\,d^3\theta\;\bm{U} \,e^{\bm{V}}\,\bm{B}
\eea
This is gauge-invariant w.r.t. the transformations
\bea\label{stuckgaugetrans}
\bm{U} \mapsto e^{\bm{\Sigma_1}}\,\bm{U}\,,\quad\quad \bm{B} \mapsto e^{\bm{\Sigma_2}}\,\bm{B}\,,\quad\quad \bm{V}\mapsto \bm{V}-\bm{\Sigma_1}-\bm{\Sigma_2}\,,
\eea
where $\bm{\Sigma_1}$ and $\bm{\Sigma_2}$ are chiral and twisted chiral respectively. To introduce the St\"uckelberg term, we define  a chiral field $\bm{S}$ and a twisted chiral field $\bm{T}$ and write down the Lagrangian\footnote{For a SUSY St\"uckelberg term in 4D Minkowski space cf.~\cite{Kors}.}  ($d^4\theta\equiv d\theta_1d\theta_2 d\thickbarsmall{\theta}_1 d\thickbarsmall{\theta}_2$)
\bea
\mathcal{L}=\left[\int\,d^3\theta\;\bm{U} \,e^{\bm{V}}\,\bm{B}-\mathrm{c.c.}\right]+{2\over \vkappa}\int\,d^4\theta\,\left(\bm{V}+\bm{S}+\bm{T}\right) \left(\thickbarsmall{\bm{V}}+\thickbarsmall{\bm{S}}+\thickbarsmall{\bm{T}}\right)
\eea
Clearly, it is gauge-invariant if one postulates the transformation rules
\bea\label{STgaugetrans}
\bm{S}\mapsto \bm{S}+\bm{\Sigma_1}\,,\quad\quad \bm{T} \mapsto \bm{T}+\bm{\Sigma_2}\,.
\eea
In fact, it is easy to see that the twisted chiral field $\bm{T}$, satisfying $D_2\bm{T}=\thickbarsmall{D}_2\bm{T}=0$, drops out of the action, since each term containing $\bm{T}$ is annihilated either by $D_2$, or by $\thickbarsmall{D}_2$, so that its top component is a total derivative. We may thus equivalently write
\bea\label{intSUSYlagr}
\mathcal{L}\simeq \left[\int\,d^3\theta\;\bm{U} \,e^{\bm{V}}\,\bm{B}-\mathrm{c.c.}\right]+{2\over \vkappa}\int\,d^4\theta\,\left(\bm{V}+\bm{S}\right) \left(\thickbarsmall{\bm{V}}+\thickbarsmall{\bm{S}}\right)
\eea
For further purposes let us write out the field $\bm{S}$ in components: 
\bea\label{Ssuperfield}
\bm{S}=s(z_-, \thickbarsmall{z}_+)+\theta_1\,\psi(z_-, \thickbarsmall{z}_+)+\thickbarsmall{\theta}_2\,\chi(z_-, \thickbarsmall{z}_+)+\theta_1\thickbarsmall{\theta}_2\,\rho(z_-, \thickbarsmall{z}_+)
\eea

As the next step we introduce an analogue of Wess-Zumino gauge. With the help of the gauge transformations~(\ref{stuckgaugetrans}), one can eliminate the components in $\bm{V}$ proportional to $1, \theta_1, \thickbarsmall{\theta}_1, \thickbarsmall{\theta}_2, \theta_1 \thickbarsmall{\theta}_1, \theta_1 \thickbarsmall{\theta}_2$. As a result, $\bm{V}$ may be cast in the form\footnote{Notice that the semi-chiral gauge superfield $\bm{V}$ does not have a $D$-term, typical of the more conventional unconstrained gauge superfield. As a result, the $D$-term constraint is also absent.}\;\footnote{Such structure of the gauge field supermultiplet  was also found in~\cite{BykovSUSY}, where the GN formulation of the $\CP^{n-1}$ sigma model was supersymmetrized without the use of superfields.}
\bea\label{WZgauge}
\bm{V}=\thickbarsmall{\theta}_1\thickbarsmall{\theta}_2\,\thickbarsmall{\mathcal{A}}(z_-, \thickbarsmall{z})+\theta_1\thickbarsmall{\theta}_1\thickbarsmall{\theta}_2\,\thickbar{\mathcal{W}}(z, \thickbarsmall{z})
\eea
In the Wess-Zumino gauge one has the residual gauge invariance $\bm{V}\mapsto \bm{V}+\thickbarsmall{\dd}\bm{\Lambda}$\, where~$\bm{\Lambda}$ is a superfield of the form~(\ref{WZgauge}):
\bea
\bm{\Lambda}=\thickbarsmall{\theta}_1\thickbarsmall{\theta}_2\,\tau(z_-, \thickbarsmall{z})+\theta_1\thickbarsmall{\theta}_1\thickbarsmall{\theta}_2\,\sigma(z, \thickbarsmall{z})
\eea
Notice that $\thickbarsmall{\dd}\bm{\Lambda}=-\{\thickbarsmall{D}_1,\thickbarsmall{D}_2 \}\bm{\Lambda}=-\thickbarsmall{D}_1\thickbarsmall{D}_2 \bm{\Lambda}-\thickbarsmall{D}_2\thickbarsmall{D}_1 \bm{\Lambda}$. Since $\bm{\Lambda}$ satisfies $D_2 \bm{\Lambda}=0$, this is a sum of chiral and twisted chiral fields. One should also perform the compensating transformation $\bm{S}\mapsto \bm{S}+\thickbarsmall{D}_1\thickbarsmall{D}_2 \bm{\Lambda}$ (the twisted chiral term $\thickbarsmall{D}_2\thickbarsmall{D}_1 \bm{\Lambda}$ drops out automatically, just like the field $\bm{T}$ above). In components, these gauge transformations~read 
\bear\label{gaugeinvcomp}
		&&\thickbarsmall{\mathcal{A}}\mapsto \thickbarsmall{\mathcal{A}}+\thickbarsmall{\dd} \tau\,,\quad\quad \thickbar{\mathcal{W}} \mapsto \thickbar{\mathcal{W}}+\thickbarsmall{\dd} \sigma\,,\\
		&&s\mapsto s- \tau\,,\quad\quad \psi \mapsto \psi- \sigma\,,
\eear
whereas the Lagrangian~(\ref{intSUSYlagr}) takes the form (up to integration by parts) 
\bear
&&\mathcal{L}\simeq \left[{1\over 2}\left(-u \, \thickbarsmall{D}v+v \,\thickbarsmall{D}u+c \,\thickbarsmall{D}b-\thickbarsmall{D}c \, b\right)-\thickbar{\mathcal{W}} bu+f \,w-d \, g-\mathrm{c.c.}\right]+
\\ \nonumber &&\quad\quad+{2\over \vkappa}\left(|\thickbarsmall{\mathcal{A}}+ \thickbarsmall{\dd}s|^2+\thickbarsmall{\chi}\,(\thickbar{\mathcal{W}}+\thickbarsmall{\dd}\psi)+\chi\,(\mathcal{W}+\dd \thickbarsmall{\psi})-\rho \thickbarsmall{\rho}\right)\,.
\eear
The fields $f, w, d, g, \rho$ are auxiliary and vanish on-shell. Clearly, $s$ plays the role of St\"uckelberg field, whereas $\psi$ is its superpartner. Using the gauge invariance~(\ref{gaugeinvcomp}), we may set
\bea
s=\psi=0\,.
\eea
Variation w.r.t. $\thickbar{\mathcal{W}}$ leads to $\thickbarsmall{\chi}=-{\vkappa\over 2}\,bu=-{\vkappa\over 2}\,\hat{J}$, whereas variation w.r.t. $\chi$ sets $\mathcal{W}=0$. Finally, variation w.r.t. $\thickbarsmall{\mathcal{A}}$ implies $\mathcal{A}={\vkappa\over 2} J$. Substituting these values back in the Lagrangian, one returns to the super-Thirring model~(\ref{SUSYThirring}).
 
 \vspace{0.3cm}
 \textbf{Remark.} An alternative strategy for the elimination of  fields is as follows. Instead of going to Wess-Zumino gauge~(\ref{WZgauge}), one could use the gauge transformations~(\ref{STgaugetrans}) to set the superfield $\bm{S}=0$ (which is the superfield analogue of $s=0$). One would then still have the residual gauge symmetry $\bm{V}\to \bm{V}-\bm{\Sigma_2}$, where $\bm{\Sigma_2}$ is twisted chiral. These residual transformations may be used to bring $\bm{V}$ to the form
\bea
\bm{V}=\thickbarsmall{\theta}_2\,\chi(z_-, \thickbarsmall{z}_+)+\theta_1\thickbarsmall{\theta}_2\,\rho(z_-, \thickbarsmall{z}_+)
+\thickbarsmall{\theta}_1\thickbarsmall{\theta}_2\,\thickbarsmall{\mathcal{A}}(z_-, \thickbarsmall{z})+\theta_1\thickbarsmall{\theta}_1\thickbarsmall{\theta}_2\,\thickbar{\mathcal{W}}(z, \thickbarsmall{z})
\eea
This is the same as $\bm{V}+\bm{S}$ in~(\ref{Ssuperfield})-(\ref{WZgauge}) above, once one sets $s=\psi=0$.

\section{The cylinder model}\label{Cylinder_CF}
	
	Classically the model~(\ref{SUSYThirring}) is equivalent to a free supersymmetric theory with target space $\mathbb{R}\times S^1$ (cylinder). To show this let us integrate out the variables $v, \thickbarsmall{v}$. This gets even more transparent if we make the change of variables $v \mapsto v- {1\over u} bc$, which eliminates the fermions from the quartic interaction. As a result, one gets the Lagrangian 
	\bear
	&&\mathcal{L}=v\thickbarsmall{\dd}u+\thickbarsmall{u}\dd \thickbarsmall{v}+b\thickbarsmall{D}c-\thickbarsmall{c} D \thickbarsmall{b}+{\vkappa\over 2} \,\big|v u\big|^2\,,\\
	&&\textrm{where}\qquad \thickbarsmall{D}c=\thickbarsmall{\dd}c-{\thickbarsmall{\dd} u\over u} c\,, \quad D\thickbarsmall{b}=\dd \thickbarsmall{b}+{\dd \thickbarsmall{u}\over \thickbarsmall{u}} \thickbarsmall{b}\,.
	\eear
One can then make the change of variables\footnote{The $B, C$-fields here are singlets and are unrelated to the $B, C$-doublets of section~\ref{Susy_Def_Sec}.}
	\bea\label{fermCOV}
	c= u \,C, \quad b= u^{-1} \,B\,,
	\eea
	which fully decouples bosonic and fermionic variables. Finally, integrating over $v, \thickbarsmall{v}$, we arrive at
	\bea
	\mathcal{L}={2\over \vkappa} \frac{|\thickbarsmall{\dd}u|^2}{|u|^2}+B\thickbarsmall{\dd}C-\thickbarsmall{C} \dd \thickbarsmall{B}\,.
	\eea
	Changing variables $u=e^{W}$ and grouping the fermionic fields into the Dirac spinor $\Psi:=\begin{pmatrix}B \\ \thickbarsmall{C}\end{pmatrix}$, this is cast in the more conventional form 
	\bea\label{freeSUSY}
	\mathcal{L}={2\over \vkappa}\, |\thickbarsmall{\dd}W|^2 + i\,\thickbarsmall{\Psi} \slashed{\dd}\Psi\,,
	\eea
	which is the free $\mathcal{N}=(2, 2)$ supersymmetric Lagrangian. 
	
	At the next step, we will show that it is possible to calculate correlation functions of the elementary fields either from~(\ref{SUSYThirring}) or from~(\ref{freeSUSY}), and the results agree. In the first approach we will use the Feynman rules of the super-Thirring model, which may be obtained from the ones in Fig.~\ref{feynfig} by simply removing all indices and Lie algebra generators. In the second, we will use the relation of the $u, v, b, c$ fields to the free fields $W, B, C$ that can be summarized as follows:
	\bea\label{vertexops}
	u=e^{W}\,,\quad\quad v=\left({2\over \vkappa}\,\dd\thickbar{W}-BC\right) \,e^{-W}\,,\quad\quad b=B\,e^{-W}\,,\quad\quad c=C\,e^{W}\,,
	\eea
together with the analogous complex conjugate expressions\footnote{In our conventions, $\thickbarsmall{v}=\left(-{2\over \vkappa}\,\thickbarsmall{\dd}W-\thickbarsmall{B}\thickbarsmall{C}\right) \,e^{-\thickbar{W}}$.}. 
	These formulas resemble the ones of bosonization, however the precise relation is unclear. Typically in bosonization one has representations in terms of vertex operators for \emph{bilinears} of elementary fields, whereas here these formulas apply to the fields themselves.
	
\begin{figure}
\centering
\begin{overpic}[abs, scale=1,unit=1mm]{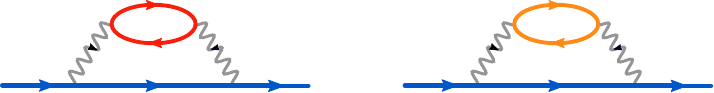}
    \put(50,15){$(a)$}
    \put(118,15){$(b)$}
\end{overpic}
\caption[Fig2loop]{Possible two-loop contributions to the 2-point function.}
\label{fig2loop}
\end{figure}

\subsection{2-point function}

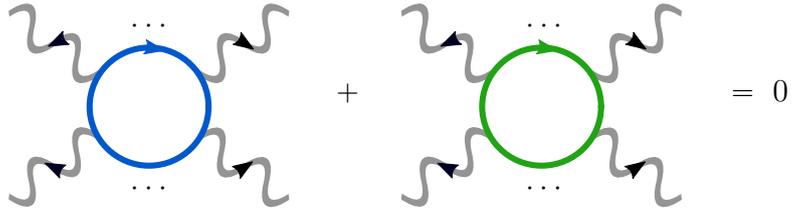
\begin{figure}
\centering
\begin{overpic}[abs, scale=0.8,unit=1mm]{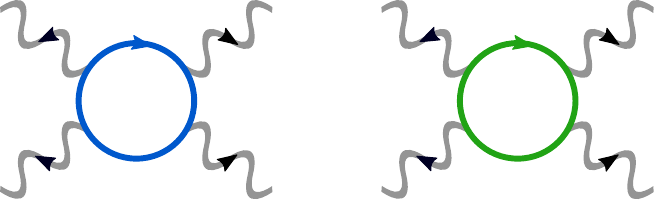}
    \put(43,14){$+$}
    \put(95,14){$=\;0$}
    \put(16,23){$\cdots$}
    \put(16,1.5){$\cdots$}
    \put(68,1.5){$\cdots$}
    \put(68,23){$\cdots$}
\end{overpic}
\caption[FigBFcancel]{Cancellation of matter loop diagrams. Same cancellation holds true between the red/orange loops as well.}
\label{figBF}
\end{figure}

Due to the structure of the Feynman rules, at one loop there are no diagrams contributing to the 2-point function $\langle u v\rangle$. First   non-trivial diagrams arise at two loops and are shown in Fig.~\ref{fig2loop}. Notice, however, that the two diagrams -- one with a bosonic loop, the other with a fermionic one -- exactly cancel each other (although each diagram in itself is non-zero: for completeness we calculate its value in Appendix~\ref{2loopcorrapp}). This can be generalized to subdiagrams with an arbitrary number of external $\mathcal{B}$-legs (see~Fig.~\ref{figBF}). It is then easy to see that any contribution to the 2-point function would involve a loop diagram of the type shown in Fig.~\ref{figBF}.  These diagrams always come in pairs with opposite signs, so that ultimately any correction to the 2-point function vanishes. Thus, the 2-point functions of elementary fields in the super-Thirring model are exactly equal to their free values.

We may then ask how this is reproduced from the calculation in the free theory~(\ref{freeSUSY}). The relevant correlation function is the one of two vertex operators shown in~(\ref{vertexops}):
\bea
\langle u(\tilde{z}_1)\, v(\tilde{z}_2)\rangle =\int\,\left({2\over \vkappa}\,\dd\thickbar{W}(\tilde{z}_2)-B(\tilde{z}_2)\,C(\tilde{z}_2)\right)\,\mathrm{exp}\left\{ -\int\!i\,dz\!\!\wedge \!\!d\bar{z}\;\mathcal{L}+W(\tilde{z}_1)-W(\tilde{z}_2)\right\}\,,
\eea
where $\mathcal{L}$ is the Lagrangian~(\ref{freeSUSY}) of the free theory. 

As we shall see, fermionic degrees of freedom are needed here simply to cancel an elementary divergence. To calculate the resulting Gaussian integral, one extremizes the action in the exponent, with the following result: 
\bea
\thickbar{W}={\vkappa\over 4\pi}\,\left(\log{|z-\tilde{z}_2|^2}-\log{|z-\tilde{z}_1|^2}\right)\,,\quad\quad W=0\,.
\eea
One easily sees that the value of the exponent on this stationary configuration is trivial. As for the operator at the front, the divergent contribution in $\dd\thickbar{W}(\tilde{z}_2)$ cancels exactly against the fermionic one, resulting in the $\vkappa$-independent answer
\bea
\langle u(\tilde{z}_1)\, v(\tilde{z}_2)\rangle ={1\over 2\pi (\tilde{z}_1-\tilde{z}_2)}\,,
\eea
in agreement with the previous analysis.

In fact, it is easy to see that there will be no corrections to any correlation functions involving only holomorphic (that is $u, v, b, c$) or only anti-holomorphic fields ($\thickbarsmall{u}, \thickbarsmall{v}, \thickbarsmall{b}, \thickbarsmall{c}$). Besides, there are elementary symmetries of the Lagrangian~(\ref{SUSYThirring}) that restrict the correlation functions. In particular, there is an obvious $\CC^\ast \times \CC^\ast$ symmetry acting as follows:
\bea
u \to \uplambda_1 u\,,\quad\quad v \to \uplambda_1^{-1} v\,,\quad\quad b \to \uplambda_2 b\,,\quad\quad c \to \uplambda_2^{-1} c\,,
\eea
and accordingly on the conjugate fields. As a result, for a correlation function to be non-zero one requires that 
\bea\label{numberoffields}
\#u=\#v, \quad \#b=\#c, \quad \#\thickbarsmall{u}=\#\thickbarsmall{v}, \quad \#\thickbarsmall{b}=\#\thickbarsmall{c},
\eea
where $\#$ means the number of insertions of the respective fields. 

\subsection{4-point function}\label{4ptsec}

The more interesting coupling-dependent correlation functions  involve both holomorphic and anti-holomorphic fields. The simplest one is the 4-point function 
\bea
\Gamma^{(4)}:=\langle u(\tilde{z}_1)\,v(\tilde{z}_2)\,\thickbarsmall{u}(z_1)\,\thickbarsmall{v}(z_2)\rangle
\eea
The only diagrams that contribute to the 4-point function are the crossed ladder diagrams. Indeed, any other diagrams would involve loops of matter fields, i.e. subdiagrams of the type shown in Fig.~\ref{figBF}, which cancel out between bosons and fermions, as discussed above.

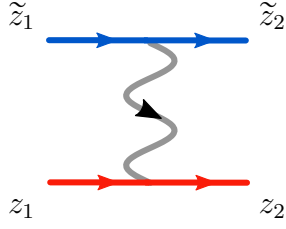
\begin{figure}
\vspace{0.5cm}
\centering
\begin{overpic}[abs, scale=1,unit=1mm]{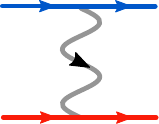}
    \put(-5,22){$ \tilde{z}_{1} $}
    \put(28,22){$ \tilde{z}_{2} $}
    \put(-5,-3){$ z_{1} $}
    \put(28,-3){$ z_{2} $}
\end{overpic}
\vspace{0.3cm}
\caption[Fig4pointtree]{Tree-level contribution to the  4-point function of bosonic fields in coordinate space.}
\label{fig4pointtree}
\end{figure}

\begin{figure}
\centering
\begin{overpic}[abs, scale=1,unit=1mm]{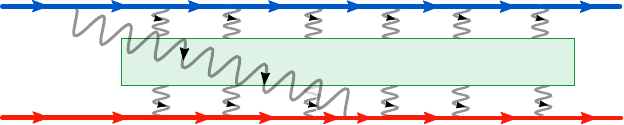}
    \put(-5,22){$ \tilde{z}_{1} $}
    \put(11,22){$w$}
    \put(25,22){$w_1$}
    \put(55,22){$\cdots$}
    \put(37,-3){$\cdots$}
    \put(77,-3){$\cdots$}
    \put(25,-3){$w_{1'}$}
    \put(50,-3){$w_{i'}$}
    \put(57,-3){$w$}
        \put(63,-3){$w_{i+1'}$}
        \put(88,-3){$w_{\ell + 1'}$}
    \put(88,22){$w_{\ell+1}$}
    \put(107,22){$ \tilde{z}_{2} $}
    \put(-5,-3){$ z_{1} $}
        \put(107,-3){$ z_{2} $}
\end{overpic}
\vspace{0.3cm}
\caption[Fig4pointtree]{Recursive relation for generating $(\ell+1)$-loop diagrams from $\ell$-loop diagrams. The green box stands for a permutation $p\in S_{\ell+1}$ determining the contraction of vertices in the top line with vertices in the lower line.}
\label{recursreldiag}
\end{figure}
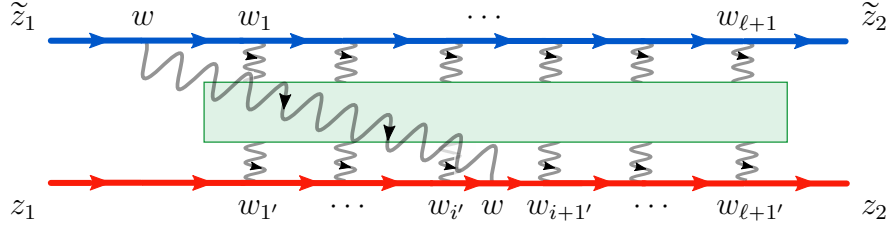

To start with, note that in free theory the value of the 4-point function is
\bea
I_{-1}(\tilde{z}_1, \tilde{z}_2 | z_1, z_2) = {1\over (2\pi)^2}\frac{1}{\tilde{z}_{12}\thickbarsmall{z}_{21}}\,.
\eea
Next, consider the tree-level diagram shown in Fig.~\ref{fig4pointtree}. Its value is\footnote{In the calculation we use the following master integral: $\int\,\frac{d^2z}{\thickbarsmall{z} (z+a)(z+b)}={\pi \over a-b}\log{\left({|a|^2\over |b|^2}\right)}$.}
\bear\label{treecontrib}
&&I_0(\tilde{z}_1, \tilde{z}_2 | z_1, z_2) = - {\vkappa\over (2\pi)^2}\,\int\,{d^2z \over (2\pi)^2} \,\frac{1}{(\tilde{z}_1-z)(z-\tilde{z}_2)(\thickbarsmall{z}_1-\thickbarsmall{z})(\thickbarsmall{z}-\thickbarsmall{z}_2)}\\ \nonumber
&&\hspace{2.5cm}= \,-{1\over (2\pi)^2}\frac{1}{\tilde{z}_{12}\thickbarsmall{z}_{21}}\times {\vkappa\over 4\pi} \log{[\mathsf{CR}(\tilde{z}_1, \tilde{z}_2| z_1, z_2)]} \,,
\eear
where $\mathsf{CR}$ is the conformal, \textit{i.e.} $\mathrm{SL}(2, \CC)$-invariant, cross-ratio
\bea\label{CR}
\mathsf{CR}(\tilde{z}_1, \tilde{z}_2| z_1, z_2)=\frac{|\tilde{z}_1-z_1|^2 |\tilde{z}_2-z_2|^2}{|\tilde{z}_2-z_1|^2 |\tilde{z}_1-z_2|^2}
\eea
The result~(\ref{treecontrib}) will be the starting point of an induction. The contribution at $\ell+1$ loops may be written as follows (here $\thickbar{w}_{0'}\equiv \thickbarsmall{z}_1$ and $\thickbar{w}_{\ell+2'}\equiv \thickbarsmall{z}_2$):
\bear\label{recursrel1}
&&\!\!\!\!\!\!\!\!\!\!\!\!\!\!\!\!\!\!\!\!\!\!I_{\ell+1}=-{\vkappa^{\ell+2}\over  (2\pi)^2}\,\sum\limits_{i=0}^{\ell + 1}\,\sum\limits_{p\in S_{\ell+1}}\!\!\int\,{d^2w \over (2\pi)^2}\,\prod\limits_{i=1}^{\ell + 1}\,{d^2 w_i\over (2\pi)^2}\,\frac{1}{(\tilde{z}_1-w)(w-w_1)\cdots (w_{\ell + 1}-\tilde{z}_2)}\times \\
\nonumber &&\hspace{2.5cm}\, \times\frac{1}{(\thickbarsmall{z}_1-\thickbar{w}_{1'})\cdots (\thickbar{w}_{i'}-\thickbar{w})(\thickbar{w}-\thickbar{w}_{i+1'})\cdots (\thickbar{w}_{\ell + 1'}-\thickbarsmall{z}_2)}\,,
\eear
where $p\in S_{\ell+1}$ is a permutation, and we have used the notation $i'=p(i)$. To prove this, we start with the $\ell$-loop contribution, which may be written as a sum over the permutations $p\in S_{\ell+1}$, each permutation generating a single crossed ladder diagram. To pass over from such an $\ell$-loop diagram to an $(\ell+1)$-loop diagram, we add a vertex, denoted $w$, in the first position in the upper line\footnote{Without loss of generality, since otherwise we could relabel the vertices, so that the new vertex is always in the first position.}  and contract it with a new vertex in \emph{some} position in the lower line (this is illustrated in Fig.~\ref{recursreldiag}). Clearly, there are $\ell+2$ such positions, and we sum over them, thus arriving at~(\ref{recursrel1}). This also gives the correct number of diagrams, since $|S_{\ell+2}|=(\ell+2)|S_{\ell+1}|$.

Now, in the second line of~(\ref{recursrel1}) we will apply the identity 
\bea\label{splitidentity}
\frac{1}{(\thickbar{w}_{i'}-\thickbar{w})(\thickbar{w}-\thickbar{w}_{i+1'})}=\frac{1}{\thickbar{w}_{i'}-\thickbar{w}_{i+1'}}\left(\frac{1}{\thickbar{w}_{i'}-\thickbar{w}}+\frac{1}{\thickbar{w}-\thickbar{w}_{i+1'}}\right)
\eea
to obtain
\begin{equation}\label{key}
		\begin{aligned}I_{\ell+1} = & -\frac{\vkappa^{\ell + 2}}{(2\pi)^2}\,\sum\limits_{p\in S_{\ell+1}}\,\int\,{d^2w\over   (2\pi)^2}\,\prod\limits_{i = 1}^{\ell + 1}\,{d^2 w_i \over  (2\pi)^2}\,\frac{1}{(\tilde{z}_1-w)\cdots (w_{\ell + 1}-\tilde{z}_2)} \\ 
		& \times \frac{1}{(\thickbarsmall{z}_1-\thickbar{w}_{1'})\cdots  (\thickbar{w}_{\ell + 1'}-\thickbarsmall{z}_2)} \sum\limits_{i=0}^{\ell+1}\,\left(\frac{1}{\thickbar{w}_{i'}-\thickbar{w}}+\frac{1}{\thickbar{w}-\thickbar{w}_{i+1'}}\right)
	\end{aligned}
\end{equation}
Most of the terms in the inner sum cancel out, and as a result one obtains the recurrence relation
\bea\label{recurr}
I_{\ell+1}(\tilde{z}_1, \tilde{z}_2 | z_1, z_2)=-\vkappa\,\int\,{d^2 w\over (2\pi)^2}\,\frac{\thickbarsmall{z}_{12}}{(\tilde{z}_1-w)(\thickbarsmall{z}_1-\thickbar{w})(\thickbar{w}-\thickbarsmall{z}_2)}\times I_{\ell}(w, \tilde{z}_2 | z_1, z_2)\,,
\eea
which is supplemented by the initial condition -- the value~(\ref{treecontrib}) of $I_0$. In Appendix~\ref{recurrsolapp} we show that this recurrence relation is solved by
\bea\label{recurrsol}
I_{\ell}(\tilde{z}_1, \tilde{z}_2 | z_1, z_2)={1\over (2\pi)^2}\frac{1}{\thickbarsmall{z}_{21} \tilde{z}_{12}}\,\frac{1}{(\ell+1)!}\,\left(-{\vkappa\over 4\pi}\right)^{\ell+1}\left[\log\,\mathsf{CR}(\tilde{z}_1, \tilde{z}_2| z_1, z_2)  \right]^{\ell+1}
\eea
The perturbation theory for the 4-point function may then be summed explicitly, resulting in the exact expression
\bea\label{Gamma4exact}
\Gamma^{(4)}=\sum\limits_{\ell=-1}^\infty I_{\ell}={1\over (2\pi)^2}\frac{1}{\thickbarsmall{z}_{21} \tilde{z}_{12}}\,\left[\mathsf{CR}(\tilde{z}_1, \tilde{z}_2| z_1, z_2) \right]^{-{\vkappa\over 4\pi}}\,.
\eea

The computation of the 4-point function on the other side of the correspondence reduces to the evaluation of the following Gaussian integral:
\begin{equation}\label{4pointfunc}
	\begin{aligned}
		& \langle u(\tilde{z}_1)\,v(\tilde{z}_2)\,\thickbarsmall{u}(z_1)\,\thickbarsmall{v}(z_2)\rangle = \\ 
		& = \int\,\left({2\over \vkappa}\,\dd \widebar{W}(\tilde{z}_2)-B(\tilde{z}_2)C(\tilde{z}_2)\right) \left(-{2\over \vkappa}\,\thickbar{\dd} W(z_2)-\thickbar{B}(z_2)\thickbar{C}(z_2)\right)\, \\ 
		& \times \exp{\left[-\,\int \!i\,dz\!\!\wedge \!\!d\bar{z}\;\mathcal{L}+W(\tilde{z}_1) - W(\tilde{z}_2) + \widebar{W}(z_1) - \widebar{W}(z_2)\right]}
	\end{aligned}
\end{equation}
Solution of the e.o.m. gives
\bea\label{4ptsol}
W={\vkappa\over 4\pi}\,\log{\left({|z-z_2|^2\over|z-z_1|^2}\right)}\,,\quad\quad \thickbarsmall{W}={\vkappa\over 4\pi}\,\log{\left({|z-\tilde{z}_2|^2 \over|z-\tilde{z}_1|^2}\right)}
\eea
Substituting these values in the action standing in the exponent, one gets $-{\vkappa\over 4\pi} \log \mathsf{CR}(\tilde{z}_1, \tilde{z}_2| z_1, z_2)$. As for the prefactor, again one sees that the role of the $BC$- and $\thickbar{B}\thickbar{C}$-fermions in the correlation function~(\ref{4pointfunc}) is to cancel the divergent contributions coming from $\dd \widebar{W}$ and $\thickbar{\dd} W$. The resulting value of the correlation function is
\bea
\langle u(\tilde{z}_1)\,v(\tilde{z}_2)\,\thickbarsmall{u}(z_1)\,\thickbarsmall{v}(z_2)\rangle={1\over (2\pi)^2}\frac{1}{\thickbarsmall{z}_{21} \tilde{z}_{12}}\,\,\left[\mathsf{CR}(\tilde{z}_1, \tilde{z}_2| z_1, z_2) \right]^{-{\vkappa\over 4\pi}}\,,
\eea
which matches the value~(\ref{Gamma4exact}) obtained by direct evaluation of the Feynman diagrams.
	
\subsection{Including fermions}\label{npointsec}

\begin{figure}
\centering
\begin{overpic}[abs, scale=1,unit=1mm]{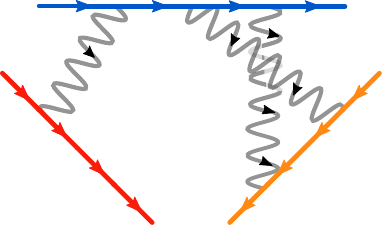}
\end{overpic}
\vspace{0.3cm}
\caption[Fig6pointtree]{Example of diagram contributing to the mixed bosonic/fermionic 6-point function.}
\label{fig6point}
\end{figure}

As discussed above, the sole role of fermions in correlation functions of bosonic operators  is to cancel bosonic loops, as in Figs.~\ref{fig2loop} and~\ref{figBF}, as well as some elementary divergent contributions. In establishing equivalence of theories~(\ref{SUSYThirring}) and~(\ref{freeSUSY}), however, we must also consider correlation functions with insertions of fermionic operators $b, c, \thickbarsmall{b}, \thickbarsmall{c}$. Let us therefore focus on the most general correlation function
\bea\label{mixedcorr}
\Gamma_{\mathrm{mixed}}:=\langle \quad\cdots\quad  \prod b(\tilde{z}_i)\,\prod c(\tilde{w}_i)\,\prod \thickbarsmall{b}(z_i)\,\prod \thickbarsmall{c}(w_i) \rangle\,,
\eea
where $\cdots$ stands for insertions of bosonic operators. The strategy is to integrate over $v, \thickbarsmall{v}$, reducing the correlation function to the one of vertex operators in the free theory~(\ref{freeSUSY}). The key difference from the correlation functions of bosonic operators is in the change of variables~(\ref{fermCOV}), which means that each insertion of $b(\tilde{z}_i)$ contributes an additional vertex operator $e^{-W(\tilde{z}_i)}$, whereas each $c(\tilde{w}_i)$ contributes $e^{W(\tilde{w}_i)}$, with similar (barred) expressions for the conjugate fermions. At the end of the day, one is effectively led to the calculation of factorised correlators of the form 
\bear\label{mixedcorrfunc}
&&\langle \quad\cdots\quad  e^{\sum W(\tilde{z}_i)-\sum W(\tilde{w}_i)+\sum \widebar{W}(z_i)-\sum \widebar{W}(w_i)} \rangle\times\\ \nonumber
&&\hspace{1.5cm}\times \;\langle \prod B(\tilde{z}_i)\,\prod C(\tilde{w}_i)\,\prod \thickbar{B}(z_i)\,\prod \thickbar{C}(w_i) \rangle_{\mathrm{free}}\,,
\eear 
where $\langle \, \dots \, \rangle_{\mathrm{free}}$ is the correlator computed in the free theory~(\ref{freeSUSY}).

Finally, what remains to compute is the correlator of the vertex operators, which is the building block of all correlation functions in the theory. Suppose one has $M$ insertions of the operators $e^W$, accordingly $M$ insertions of $e^{-W}$ as required by~(\ref{numberoffields}), and $\thickbar{M}$ insertions of the conjugate operators (one may have $\thickbar{M}\neq M$). The relevant classical solution is a simple generalization of~(\ref{4ptsol}):
\bea\label{Mptsol}
W={\vkappa\over 4\pi}\,\sum\limits_{i=1}^{\thickbar{M}}\,\log{\left({|z-w_i|^2\over|z-z_i|^2}\right)}\,,\quad\quad \widebar{W}={\vkappa\over 4\pi}\,\sum\limits_{i=1}^{M}\,\log{\left({|z-\tilde{w}_i|^2 \over|z-\tilde{z}_i|^2}\right)}
\eea
The value of the correlation function of the vertex operators is a product of Koba-Nielsen factors:  
\bear
\langle \;e^{\sum_{i=1}^M \left(W(\tilde{z}_i)- W(\tilde{w}_i)\right)+\sum_{i=1}^{\thickbar{M}} \left(\widebar{W}(z_i) - \widebar{W}(w_i)\right)}\; \rangle=
\prod\limits_{i=1}^M\;\prod\limits_{j=1}^{\thickbar{M}}\;\left({|z_j-\tilde{w}_i|^2 |w_j-\tilde{z}_i|^2\over|z_j-\tilde{z}_i|^2 |w_j-\tilde{w}_i|^2}\right)^{{\vkappa\over 4\pi}}
\eear 

In the full correlation function~(\ref{mixedcorr}) one should also take into account the prefactors entering the vertex operators~(\ref{vertexops}), similarly to the way we did for the 2- and 4-point functions. We believe that it is also possible to rederive the final expressions for the correlation functions starting directly from the (generalized) crossed ladder diagrams of the super-Thirring model, generalizing the method described above in the case of the 4-point function. An example of such diagram for a higher-point correlation function is shown in Fig.~\ref{fig6point}.

\newpage
\section{Conclusions and outlook}\label{conclusionsec} 

In the present work we have further developed the Gross-Neveu-based approach \cite{BykovGN} to sigma models with various target spaces. Specifically, we have shown that the GN formulation may be applied to the deformed supersymmetric sigma model with target space $ \bb{CP}^{1}$. This proposal not only provides a supersymmetric extension of the Fateev-Onofri-Zamolodchikov \cite{FOZ} model, but also allows using various field theoretic techniques to investigate its properties. 

We have shown that, although the GN model is not manifestly supersymmetric, its hidden $ \cl{N} = (2,2) $ supersymmetry relies on a certain identity that holds for the classical $\mathfrak{sl}_2$ $r$-matrix. It is an important task for the future to extend the GN formulation to deformed $\CP^{n-1}$ models, since for $n>2$ the analogous relation for $\mathfrak{sl}_n$ $r$-matrices no longer holds, and our construction is not directly applicable. In a more geometric language, this means that one would need to extend our framework to include generalized Kähler target spaces. Another, perhaps related, goal would be to make the GN formulations manifestly supersymmetric by writing the corresponding Lagrangians in superspace. In the present paper we have been able to do this for the case of the super-Thirring model, which should be thought of as the \emph{abelian} GN model. Already this simplest example involves interesting combinations of chiral, twisted and semi-chiral fields that may draw hopes for further non-Abelian generalizations. 

Equivalence of the superdeformed GN model with the superdeformed $ \bb{CP}^{1} $ sigma model has been established by passing over to inhomogeneous coordinates. In the course of this the deformed $ \bb{CP}^{1} $ metric and the Riemann tensor entering the four-fermion interaction arise automatically from the GN model upon integration over auxiliary variables, leading to the result that is fully consistent with standard approaches (cf.~\cite{MirrorBook}).

We have computed the $ \beta $-function of the supersymmetric deformed GN model at one and two loops. The one-loop result agrees with an analogous one for the purely bosonic model of~\cite{FOZ} whereas we find no correction at two loops. This is consistent with the recent results of~\cite{Gamayun}, as well as with the older conjecture~\cite{LeClair}.  A particularly important task would be to study higher-loop corrections to the $ \beta $-function in these supersymmetric models, since it was observed in~\cite{BykovBeta}  that, at four loops, a characteristic regularization scheme dependence arises\footnote{At first sight this is qualitatively compatible with old results on $\beta$-functions of K\"ahler sigma models~\cite{Grisaru1, Grisaru2}. On the other hand, for K\"ahler homogeneous target spaces the one-loop expression is believed to be exact~\cite{MPS}. It has been shown recently that this might also be true even in deformed cases when the corresponding model is integrable~\cite{AlfimovLitvinov}.}.

Our superdeformed construction admits particularly interesting conformal (UV) limits. It turns out that our model reduces to the \textit{supercigar}~\cite{JackJonesCigar, Hori:2001ax, KarchTong} and \textit{supercylinder} models under special parametric limits  
\begin{equation}\nonumber
	\tx{Superdeformed } \bb{CP}^{1} \quad \mapsto \quad 
	\begin{cases}
		\tx{Supercigar: } u\rightarrow \fk{s}^{\frac{1}{4}} u,\, \fk{s}\rightarrow 0 \\
		\tx{Supercylinder } \mathbb{R} \times S^{1}:\, \fk{s}\rightarrow 0 \\
	\end{cases}
\end{equation}
{\hspace{3.0cm} Conformal limits of the SUSY sausage model.} \\ [1ex] 
These limiting cases have consistent dual descriptions in the GN formalism. As an example of this, we have provided a relation between the supercylinder and super-Thirring models. It should be noted that the map between operators in the two models is nontrivial and somewhat akin to bosonization rules. In the super-Thirring model we have been able to obtain an exact answer for the 4-point function of elementary fields by explicitly summing all relevant Feynman diagrams. We then found full agreement with a vertex operator computation in the $ \cl{N} = (2,2) $ cylinder sigma model.    

A possible direction for the future involves the study of mirror duals of the proposed supersymmetric models, which are known to be perturbed Landau-Ginzburg theories~(cf.~\cite{CecottiVafa, Fendley:1992dm, Fendley:1993pi}). Perhaps most promising in this regard is the relation between the supercigar model and $ \cl{N} = 2 $ Liouville theory~\cite{Hori:2001ax}. We leave all these questions for future work.

\vspace{0.5cm}
\nd \textbf{Acknowledgments.} We would like to thank  C.~Ahn, Z.~Bajnok, F.~Hassler, B.~Hoare, N.~Ishtiaque, E.~Ivanov, C.~Klimčík, M.~Kontsevich, G.~Korchemsky, V.~Krivorol, S.~Lacroix, A.~Losev, A.~Smilga, A.~Tseytlin and B.~Vicedo for discussions, and especially P.~Fendley and M.~Roček for comments on the manuscript. This work has been supported by Russian Science Foundation grant \href{https://rscf.ru/en/project/22-72-10122/}{RSCF-22-72-10122}. D.B. would like to thank the Institut des Hautes \'Etudes Scientifiques, where part of this work was done, for hospitality.

\newpage 

	\appendix

\makeatletter
\renewcommand\section{\@startsection {section}{1}{\z@}
	{-3.5ex \@plus -1ex \@minus -.2ex}
	{2.3ex \@plus.2ex}
	{\normalfont\large\bfseries }}
\renewcommand\subsection{\@startsection{subsection}{2}{\z@}
	{-3.25ex\@plus -1ex \@minus -.2ex}
	{1.5ex \@plus.2ex}
	{\normalfont\normalsize\bfseries}}
\makeatother
	
	\begin{center}
		\Large \textbf{Appendix}
	\end{center}
	
	\section{Eliminating SUSY auxiliary fields}\label{auxfieldsapp}
	
	In this Appendix we show how one can eliminate auxiliary fields from the supersymmetric Lagrangian~(\ref{11SUSYlagr})
	\bea
	\mathcal{L}=-\int\,d^2\theta\, \left(\bm{B}  \thickbar{D}  \bm{U}+\thickbar{\bm{U}}D \thickbar{\bm{B}}+{\vkappa \over 2}\,  \bm{J} \thickbar{\bm{J}} \right)\,,
	\eea
	First let us write out those terms in the kinetic piece of the Lagrangian that contain auxiliary fields (see the decomposition~(\ref{UBdecomp})):
	\bea
	\mathcal{L}_0^{\mathrm{aux}}=\int\,d\thickbar{\theta}\, \thickbar{U}_1 \thickbar{B}_1-\int\,d\theta\,B_1 U_1\,.
	\eea
	Here the field $B_1$ is bosonic (commutative), whereas $U_1$ is Grassmann. Expanding in~$\theta$, we get $B_1=B_{10}+\theta\,B_{11}$ and $U_1=U_{10}+\theta\,U_{11}$, so that
	\bea\label{L0aux}
	\mathcal{L}_0^{\mathrm{aux}}=\thickbar{B}_{10}\thickbar{U}_{11}+\thickbar{B}_{11}\thickbar{U}_{10}-B_{11}U_{10}-B_{10}U_{11}
	\eea
	To write out the interaction term, first we introduce the current superfield

\begin{equation}
	\bm{J}=\bm{U}\bm{B}=J_0+\thickbar{\theta}\,J_1 \, ,
\end{equation}
where 
\begin{equation}
	\begin{aligned}
		&J_0=\hat{J}+\theta J \\ 
		&J_1=J_{10}+\theta J_{11}
	\end{aligned}
	\qquad
	\begin{aligned}
		& J_{10}=U_{10}b+uB_{10} \\ 
		& J_{11}=U_{11}b+U_{10}v+cB_{10}+uB_{11}
	\end{aligned}
\end{equation} 
The interaction Lagrangian in these terms becomes 
	\bea
	\mathcal{L}_{\mathrm{int}}=-\int\,d^2\theta\,\bm{J} \thickbar{\bm{J}}=J\thickbarsmall{J}-J_{10}\thickbar{J}_{10}-J_{11}\hat{\thickbarsmall{J}}+\hat{J} \thickbar{J}_{11}\,.
	\eea
	Combining this with~(\ref{L0aux}), we may write the part of the full Lagrangian depending on auxiliary fields:
	\bear
	&&\mathcal{L}^{\mathrm{aux}}=\thickbar{B}_{10}\thickbar{U}_{11}+\thickbar{B}_{11}\thickbar{U}_{10}-B_{11}U_{10}-B_{10}U_{11}-{\vkappa \over 2}\big|U_{10} b+u B_{10}\big|^2-\\ \nonumber
	&&-{\vkappa \over 2}\left(U_{11}b+U_{10}v+cB_{10}+uB_{11}\right)\hat{\thickbarsmall{J}}+{\vkappa \over 2}\hat{J}\left(\thickbar{U}_{11}\thickbarsmall{b}+\thickbar{U}_{10}\thickbarsmall{v}+\thickbarsmall{c}\thickbar{B}_{10}+\thickbarsmall{u}\thickbar{B}_{11}\right)
	\eear
	Notice that $B_{11}, U_{11}, \thickbar{B}_{11},  \thickbar{U}_{11}$ enter as Lagrange multipliers. Varying w.r.t. these fields, one gets
	\bear
	&&B_{11}:\quad\quad U_{10}=-{\vkappa \over 2} u \hat{\thickbarsmall{J}}\,,\quad\quad \thickbar{B}_{11}:\quad\quad \thickbar{U}_{10}={\vkappa \over 2} \hat{J}\thickbarsmall{u}\\
	&&U_{11}:\quad\quad B_{10}=-{\vkappa \over 2} b\hat{\thickbarsmall{J}}\,,\quad\quad \thickbar{U}_{11}:\quad\quad \thickbar{B}_{10}=-{\vkappa \over 2} \hat{J}\thickbarsmall{b}
	\eear
	Varying w.r.t. the remaining fields, we find
	\bear
	&&U_{10}:\quad\quad B_{11}={\vkappa \over 2} v \hat{\thickbarsmall{J}}\,,\quad\quad \thickbar{U}_{10}:\quad\quad \thickbar{B}_{11}=- {\vkappa \over 2}\hat{J}\thickbarsmall{v}\\
	&&B_{10}:\quad\quad U_{11}=-{\vkappa \over 2} c \hat{\thickbarsmall{J}}\,,\quad\quad \thickbar{B}_{10}:\quad\quad \thickbar{U}_{11}=-{\vkappa \over 2}\hat{J} \thickbarsmall{c}
	\eear
	Substituting these values in the above Lagrangian one finds that $\mathcal{L}^{\mathrm{aux}}=0$ on-shell. 
	
	Recall the SUSY variation~(\ref{N11SUSYvar}). Taking its complex conjugate, one finds $\delta \thickbar{U}_0=\thickbarsmall{\epsilon}\, \thickbar{Q}\thickbar{U}_0+\epsilon \thickbar{U}_1$, implying
	\bea
	\delta \thickbarsmall{u}=\thickbarsmall{\epsilon}\,\thickbarsmall{c}+\epsilon \thickbar{U}_{10}=\thickbarsmall{\epsilon}\,\thickbarsmall{c}+{\vkappa \over 2} \epsilon\hat{J}\thickbarsmall{u} \quad\quad \textrm{and}\quad\quad \delta \thickbarsmall{c}=-\thickbarsmall{\epsilon} \,\thickbar{\dd} \thickbarsmall{u}-\epsilon \thickbar{U}_{11}=-\thickbarsmall{\epsilon}\, \thickbar{\dd} \thickbarsmall{u}+{\vkappa \over 2} \epsilon \hat{J} \thickbarsmall{c}\,,
	\eea
	which matches the transformation rules~(\ref{compenstrans}) found earlier. One can analogously compare the transformation rules of the remaining fields. The conclusion is that the auxiliary fields are necessary in order to reproduce the compensating transformations.
	
	\section{$O(\vkappa^2)$ correction to the 2-point function}\label{2loopcorrapp}
	
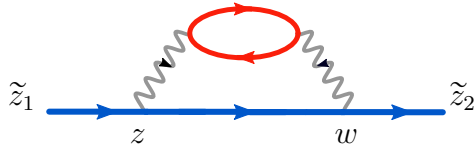
\begin{figure}
\centering 
\centering
\begin{overpic}[abs, scale=1,unit=1mm]{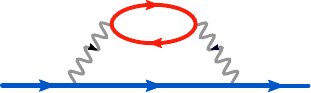}
    \put(-5,2){$\tilde{z}_1$}
    \put(53,2){$\tilde{z}_2$}
    \put(11,-3){$z$}
    \put(38,-3){$w$}
    \end{overpic}
    \vspace{0.2cm}
\caption[Fig2loopcalc]{Two-loop contribution to the 2-point function in the pure bosonic/fermionic Thirring model.}
\label{fig2loopcalc}
\end{figure}

	In this Appendix we compute the correction to the 2-point function for models where full cancellations between bosons and fermions are absent (for level $k\neq 0$, in the language of~\cite{BykovBeta}). Such are the pure fermionic or pure bosonic Thirring models, for example. The relevant diagram is shown in Fig.~\ref{fig2loopcalc} and gives the integral
	\bea
	I_1=\vkappa^2\,\int\,{d^2z\over (2\pi)^2}\,{d^2w\over (2\pi)^2}\,\frac{1}{(\tilde{z}_1-z)(z-w)(w-\tilde{z}_2)(\thickbarsmall{z}-\thickbar{w})^2}
	\eea
	We split the ratio $1\over (z-w)(w-\tilde{z}_2)$ using the identity~(\ref{splitidentity}) to obtain:
	\bea
	I_1=\vkappa^2\,\int\,{d^2z\over (2\pi)^2}\,\frac{1}{(\tilde{z}_1-z)(z-\tilde{z}_2)}\,\int\,{d^2w\over (2\pi)^2}\,\left({1\over z-w}+{1\over w-\tilde{z}_2}\right)\,\frac{1}{(\thickbarsmall{z}-\thickbar{w})^2}
	\eea
	In the principal value prescription, the first term in the second integral is zero, whereas for the second term we use $\int {d^2w \over (2\pi)^2}\,\frac{1}{(w-a)(\thickbarsmall{b}-\thickbar{w})^2}=\frac{1}{4\pi(\thickbar{a}-\thickbarsmall{b})}$. As a result, one is left with a single integral
	\bea
	I_1={\vkappa^2\over 4\pi}\,\int\,{d^2z\over (2\pi)^2}\,\frac{1}{(z-\tilde{z}_1)|z-\tilde{z}_2|^2}=\left({\vkappa \over 4\pi}\right)^2\frac{1}{\tilde{z}_2-\tilde{z}_1}\,\log{\left({|\tilde{z}_2-\tilde{z}_1|^2\over \varepsilon^2}\right)}\,,
	\eea
	where $\varepsilon$ is a UV cutoff. This is compatible with the exact result for the 2-point function found in~\cite{Freedman2}.

	\section{Solving the recurrence relation for the 4-point function}\label{recurrsolapp}
	
	Here we will show that the recurrence relation~(\ref{recurr}) is solved by~(\ref{recurrsol}). To this end let us evaluate the integral in~(\ref{recurr}) upon the substitution of the ansatz~(\ref{recurrsol}):
	\begin{equation}\label{key}
		I_{\ell+1}(\tilde{z}_1, \tilde{z}_2 | z_1, z_2)={-1\over (2\pi)^2}\left(-{\vkappa \over 4\pi}\right)^{\ell+2}\frac{1}{(\ell+1)!}\, \int\,{d^2 w\over \pi}\,\frac{\left[\log \mathsf{CR}(w, \tilde{z}_2| z_1, z_2)  \right]^{\ell+1}}{(\tilde{z}_1-w)(w-\tilde{z}_2)(\thickbarsmall{z}_1-\thickbar{w})(\thickbar{w}-\thickbarsmall{z}_2)}
	\end{equation}
	To evaluate this integral, first one should observe that the quantity $\thickbarsmall{z}_{21}\tilde{z}_{12} I_{\ell+1}(\tilde{z}_1, \tilde{z}_2 | z_1, z_2)$ is $\mathrm{SL}(2, \CC)$-invariant. Thus, one may write
	\bea\label{Il1}
	I_{\ell+1}(\tilde{z}_1, \tilde{z}_2 | z_1, z_2)={-1\over (2\pi)^2}\left(-{\vkappa \over 4\pi}\right)^{\ell+2}\frac{1}{(\ell+1)!}\,\frac{1}{\thickbarsmall{z}_{21}\tilde{z}_{12}}\,\mathbf{F}(\mathsf{CR}(w, \tilde{z}_2| z_1, z_2))\,,
	\eea
	where $\mathbf{F}(x)$ is a function of the conformal cross-ratio. To find this function, we may set three points to convenient values: let us send $ z_2\to \infty, z_1\to 0, \tilde{z}_2\to 1$. From~(\ref{CR}) we see that in this limit the cross-ratio is simply $\mathsf{CR}(w, \tilde{z}_2| z_1, z_2)\mapsto |w|^2$, so that
	\bear
	&&\mathbf{F}(\mathsf{CR}(z, 1 | 0, \infty))=\\ \nonumber
	&&=(z-1)\,\int\,{d^2 w\over \pi}\frac{1}{\thickbar{w}(z-w)(w-1)}\times (\log{|w|^2})^{\ell+1}=-\frac{1}{\ell+2}\,(\log{|z|^2})^{\ell+2}
	\eear
	Substituting this into~(\ref{Il1}), we obtain agreement with~(\ref{recurrsol}), thus completing the proof.

\setstretch{0.8}
\setlength\bibitemsep{5pt}
\printbibliography

\end{document}